%% file: paper.tex
\documentclass[12pt, letterpaper]{JHEP3}

\usepackage{amssymb, amsmath, amsopn, amsthm}

\usepackage{epsfig}

\usepackage{psfrag}

\usepackage{subfigure}

\input texpref.tex

\title{Towards M2-brane Theories for Generic Toric Singularities}

\author{Sebasti\'an Franco$^{1,2}$, Amihay Hanany$^{3}$, Jaemo Park$^{4,5,6}$ and Diego Rodr\'iguez-G\'omez$^{1}$

\\

~\\

$^1$Joseph Henry Laboratories, Princeton University \\
Princeton, NJ 08544, USA \\ \vspace{0.3cm}

$^2$KITP, University of California \\
Santa Barbara, CA 93106-4030, USA \\ \vspace{0.3cm}

$^3$Theoretical Physics Group, The Blackett Laboratory, Imperial College London, \\ 
Prince Consort Road, London, SW7 2AZ, U.K.\\ \vspace{0.3cm}

$^4$Department of Physics, POSTECH \\
Pohang 790-784, Korea \\ \vspace{0.3cm}

$^5$Postech Center for Theoretical Physics (PCTP), POSTECH \\ 
Pohang 790-784, Korea \\ \vspace{0.3cm}

$^6$Department of Physics, Stanford University \\
Stanford, CA 94305-4060, USA \\ \vspace{0.8cm}

\email{sfranco@kitp.ucsb.edu,a.hanany@imperial.ac.uk}\\
\email{jaemo@postech.ac.kr,drodrigu@princeton.edu}\\

}

\abstract{We construct several examples of $(2+1)$ dimensional ${\cal N}=2$ supersymmetric Chern-Simons theories, whose moduli space is given by non-compact toric
Calabi-Yau four-folds, which are not derivable from any
(3+1) dimensional CFT. One such example is the gauge theory
associated with the cone over $Q^{111}$. For several examples, we
explicitly confirm the matter content, superpotential interactions and RG flows suggested by crystal models. Our results provide additional support to the idea that crystal models are relevant for describing the structure of these CFTs. 
}

\preprint{PUPT-2279 \\ NSF-KITP-08-125 \\ Imperial/TP/08/AH/08 \\SU-ITP-08/22}

\def\be{\begin{equation}}

\def\ee{\end{equation}}

\def\bea{\begin{eqnarray}}

\def\eea{\end{eqnarray}}

\def\l{\lambda}
\def\m{\mu}
\def\n{\nu}
\def\nn{\nonumber}

\newcommand{\IIC}{\mathbb{C}}
\newcommand{\IIZ}{\mathbb{Z}}

\begin{document}

\tableofcontents


\section{Introduction}

Until recently, the AdS$_4$/CFT$_3$ correspondence was underdeveloped compared with its cousin
AdS$_5$/CFT$_4$ \cite{Maldacena:1997re,Gubser:1998bc,Witten:1998qj}. One of the main reasons for that was our lack of understanding of 
CFT$_3$ underlying M2 brane theories probing a Calabi-Yau
four-fold. M2-brane theories remained elusive compared with the theories
of D3-branes, which can be studied by the usual open string analysis. 
Even the simplest CFT$_3$ theory corresponding to M2-branes probing
$\IIC^4$ seemed to pose great difficulty. Since it has ${\cal N}=8$
supersymmetry
and apparently it was difficult to go beyond ${\cal N}=3$ supersymmetry in
a (2+1)d field theory, the M2 theory was regarded as some unknown field
theory yet to be constructed. Recently, the status quo has changed drastically. One of the key observations
was made by Schwarz \cite{sch}. He suggested that by introducing Chern-Simons terms and turning off the gauge kinetic terms in a
suitable limit, one can construct (2+1)d field theories with more
than ${\cal N}=3$ supersymmetry. One avatar of such idea, in retrospect, was constructed by Bagger and Lambert \cite{BL1, BL2, BL3} and independently by Gustavsson \cite{gus1, gus2}. 
At first sight, a key role was played by 3-algebras, which do not have a usual field theory structure. Subsequently, it was shown that the theory can be recast as an ordinary field theory \cite{rams}.
Since then, we have rapidly piled up various higher supersymmetric
theories of Chern-Simons theories with ${\cal N}=6$, $5$, $4$ and $3$
\cite{ABJM,gw,hl3p, hl45,Benna:2008zy,Schnabl,Berg, Bagger3,Jaff}, increasing the number of understandable AdS$_4$/CFT$_3$ pairs.
In particular, the conjectured CFT$_3$ dual of coincident M2-branes probing $\IIC^4$ is the ${\cal N}=6$ CS theory worked out by Aharony, Bergman, Jafferis and Maldacena (ABJM). Various checks of this proposal have been performed, such as computing the moduli space, superconformal index \cite{Bhatta} and higher order interactions due to instantons \cite{Hosomichi:2008ip}. ${\cal N}=4$ theories for M2-branes on $(\IIC^2/\IIZ_n)^2$ and orientifolds thereof were constructed in \cite{hl3p}.\footnote{There are other ${\cal N}=4$ theories associated with $\IIC^2/\IIZ_n \times \IIC^2/\IIZ_m$, with $ n \neq m$, which involve auxiliary fields \cite{Imamura}.}

In the case of D3-branes, the AdS$_5$/CFT$_4$ correspondence has a very rich structure of theories with ${\cal N}=1$ supersymmetry (i.e. four
supercharges) \cite{Hanany1, Hanany2, Hanany3, Hanany4}. 
The most comprehensive class of dual pairs has been achieved for D3-branes probing non-compact toric Calabi-Yau 3-folds. Given this situation, one might wonder if similar structures have yet to be discovered in AdS$_4$/CFT$_3$ with ${\cal N}=2$ supersymmetry (namely four supercharges). Several authors have already initiated such study \cite{Martelli, Ami, Ueda, Imamura2}. So far the field theory constructions have been restricted to
those derived from (3+1)d theories, i.e. theories with the same quiver diagrams and superpotentials as those of (3+1)d. This represents considerable progress, but this set of theories is far from generic. One should go beyond this approach to attack general ${\cal N}=2$ AdS$_4$/CFT$_3$. Here we initiate such 
study. In the current paper, we construct several examples of the
theories
which cannot come from the (3+1)d quiver theories. One famous example is the theory of M2-branes probing the cone over $Q^{111}$, which is expected to have a sextic superpotential. This theory is the
(2+1)d analogue of the theory of D3-branes probing the conifold worked out by Klebanov and Witten \cite{Klebanov}. Using the recently developed formalism of \cite{Ami} and partly guided by crystal models \cite{Lee:2006hw, Lee:2007kv, Kim:2007ic}, 
we explicitly construct a (2+1)d theory whose moduli space is indeed $C(Q^{111})$, where $C(M)$ denotes
the cone over the manifold $M$.\footnote{We will use this notation to denote both real and complex cones over certain manifolds. We are confident the difference will be clear in each specific case.} In addition, we work out several other theories that cannot have a (3+1)d origin. 
Obviously, our constructions just touch the tip of
an iceberg and an extensive investigation of all related issues 
is beyond the scope of the current paper. 
One feature worth mentioning, though, is that there can be more than one theory with the same moduli space. This is reminiscent of Seiberg 
duality or toric duality \cite{Feng2,Feng1} in (3+1)d. Apparently, there are far more possibilities in (2+1)d than in (3+1)d. 

The paper is organized as follows.
In section 2, we briefly summarize some aspects of ${\cal N}=2$ Chern-Simons theories with matter and crystal models needed for later sections. In section \ref{section_3d_not_4d}, we mention some features of (2+1)d theories derivable from (3+1)d and try to characterize them.
In section \ref{section_Q111}, we construct the theory whose moduli space is
$C(Q^{111})$. In section \ref{section_dP3}, we construct the theory for $C(dP_3) \times \IIC$ and show that, upon addition of masses for adjoint fields, it is connected to that of $C(Q^{111})$ by an RG-flow. This RG flow is suggested by crystal models. In section \ref{section_other_pair}, we construct another pair of
models related by a similar RG-flow, the theories for $D_3$ and $\IIC^3/(\IIZ_2 \times \IIZ_2) \times \IIC$. Interestingly, the theories in section \ref{section_other_pair} are related to those in sections \ref{section_Q111}  and \ref{section_dP3} by a simple flip in the charges of some matter fields. In section \ref{section_ZN_ZN}, we propose the CS theories for $\IIC^3/(\IIZ_n \times \IIZ_n) \times
\IIC$. This proposal is explicitly checked for $n=3$ in appendix A. 
Section \ref{section_partial_resolution} discusses partial resolution and how it connects the theories we have studied. We conclude in section \ref{section_conclusions}. In appendix B, we present some thoughts about parity invariance in these models.


\section{Preliminaries}

\subsection{Moduli spaces of (2+1)d CS theories}

The moduli space of the theories of our interest can be
computed following \cite{Ami}. We now summarize the procedure. A (2+1)d ${\cal N}=2$ Chern-Simons(CS) theory with bifundamental and adjoint matter is given, in ${\cal N}=2$ superspace notation, by the following Lagrangian 
\beq
{\rm Tr}\left( -\int d^4\theta \sum_{X_{ab}} X_{ab}^\dagger e^{-V_a}
  X_{ab} e^{V_b}  - i  
\sum_a k_a\int_0^1 dt  V_a \bar D^\alpha (e^{t V_a}D_\alpha e^{-t V_a}) 
 +  \int d^2\theta  W(X_{ab}) + {\rm c.c.}\right) ~,
\eeq
where $V_a$ are vector supermultiplets and $X_{ab}$ denote chiral supermultiplets transforming in the fundamental representation of gauge group $a$ and the anti-fundamental representation of gauge group $b$. For $a=b$, this corresponds to adjoint matter for gauge group $a$. We take $\sum k_a=0$. This is a necessary condition for the moduli space to be four complex dimensional.
Recall that in 2+1 dimensions a vector superfield has the expansion
\beq
V= -2 i \theta\bar \theta \sigma + 2 \theta \gamma^\mu \bar\theta
A_\mu + \cdot\cdot\cdot + 
\theta^2\bar\theta^2 D ~,
\eeq
where we omitted the fermionic part. Compared to 3+1 dimensions, there is a new scalar field $\sigma$. We can write all terms contributing to
the scalar potential in the Lagrangian
\beq
{\rm Tr} \left(- 4 \sum_a  k_a \sigma_a D_a + \sum_a  D_a \mu_a(X)
  -\sum_{X_{ab}} (\sigma _a 
X_{ab} - X_{ab} \sigma_b)(\sigma _a X_{ab} - X_{ab} \sigma_b)^\dagger  
-\sum_{X_{ab}} |\partial_{X_{ab}} W|^2\right) ~. 
\eeq
$\mu_a(X)$ is the moment map for the $a$-th gauge group
\beq  
\mu_a(X)= \sum_b X_{ab} X_{ab}^\dagger - \sum_c X_{ca}^\dagger X_{ca} + [X_{aa},X_{aa}^\dagger] ~,
\eeq
and gives the D-term. Here we use the same terminology of (3+1)d. 

By integrating out the auxiliary fields $D_a$, we see that the bosonic
potential is a sum of squares. 
The vacua can be found by looking for vanishing of the scalar potential. This gives rise to a set of matrix equations
\bea
\partial_{X_{ab}} W &=& 0\nonumber\\
 \mu_a(X) &=& 4 k_a \sigma_a  \nonumber\\
\sigma _a X_{ab} - X_{ab} \sigma_b &=& 0
\eea
The solutions to these equations automatically satisfy $D_a=0$ and correspond to supersymmetric vacua. F-term constraints are exactly 
as in the (3+1)d case, while D-term constraints are modified. The solution to only the F-terms is a useful object called the master space \cite{Forcella:2008bb}, which is also a toric variety.

Let us consider the abelian case. The supersymmetric conditions
set all $\sigma_a$ equal to a given 
value $\sigma$.
The remaining equations 
\beq
\mu_a(X) = 4 k_a \sigma
\eeq
look like standard D-term equations with a set of effective FI terms $\zeta_a=4k_a \sigma$. 
Since $\sum_a k_a=0$, one of these equations is redundant. Call $G$ the number of
gauge groups. We are left with $G-1$ equations. By taking integer linear combinations
of the equations, we can set $G-2$ equations to the form
\beq
\tilde{\mu}_i (X) =0 \, , \qquad\qquad i=1,...G-2
\eeq
where the index $i$ identifies $G-2$ linear combinations of the gauge
group, orthogonal to the 
direction determined by the FI parameters $\zeta_a$. These combinations are easily identified as the kernel of the following matrix \cite{Hanany:2008fj}

\beq
C=\left(\begin{array}{ccccc}

\ 1 \ & \ 1 \ & \ \ldots \ & \ 1 \ & \ 1 \ \\ 
k_1 & k_2 & \ldots & k_{G-1} & k_{G}
\end{array}\right) ~.
\eeq
We see that we are imposing the vanishing of the D-terms for $G-2$ $U(1)$ gauge groups. As usual, combining D-term constraints with U(1) gauge transformations is equivalent to modding out by the complexified gauge group. The equation for the remaining $U(1)$ gauge field looks like a D-term condition with a FI term. 
However, it does not add further constraints: 
it simply determines the value of the auxiliary field $\sigma$. 
Analogously we do not need to mod out by the remaining $U(1)$ gauge group.
As explained in detail in \cite{tong,ABJM}, the $U(1)$ is coupled to the
overall $U(1)$ gauge field by the Chern-Simons coupling and leaves
a discrete symmetry $\IIZ_k$, where $k=\gcd (\{k_a\})$. Note that since we are obtaining the 4-complex dimensional moduli space by imposing $G-2$ D-terms, the master space is $G+2$ dimensional.

The moduli space is non-compact CY$_4$-fold and is interpreted as the transverse space to one M2-brane in M-theory probing such geometry. In the non-abelian case, the moduli space is the symmetric product of $N$ copies of the abelian moduli space \cite{Ami}.

The computation of the moduli space closely resembles
the (3+1)d case, with a simple modification concerning D-terms.
In practice, we will use the machinery of toric geometry.
We refer the reader to \cite{Feng2}, to which notation we adhere, for 
a comprehensive review of its application to this problem.

We close this section with a comment on two classes of models: those in which 
all $k_i\neq 0$ and those in which some $k_i=0$. For the first class, all vector 
multiplets become massive and hence it is easier to think about their IR limit.
On the other hand, we can also argue that there is no objection to considering models in
the second class. When computing the moduli space, we quotient by the $U(1)$ gauge group(s)
with $k_i=0$. Hence, we only consider variables that are invariant under these gauge group(s).
We can think about the corresponding gauge fields as auxiliary fields (Lagrange multipliers).
For all the quivers in this paper, we find CS levels in both classes that reproduce the desired
geometries. Since both of them seem to work at the level accessible 
at the moment, we list both possibilities.

\subsection{Minimal review of crystal models}

We now give a brief review of the M-theory crystal models \cite{Lee:2006hw,Lee:2007kv, Kim:2007ic}. 
Crystal models relate a toric CY$_4$ to a three-dimensional periodic 
graph (crystal). This crystal is conjectured to encode information about the CFT$_3$ on M2-branes probing the corresponding CY$_4$. 

\begin{figure}[htbp]
\begin{center}
\includegraphics[width=5cm]{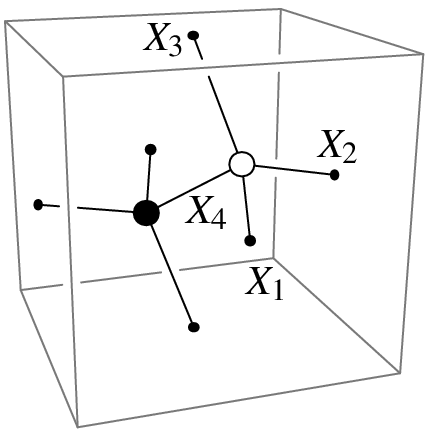}
\caption{Crystal of $\IIC^4$ adopted from \cite{Lee:2007kv}.} \label{C4}
\end{center}
\end{figure}

The toric diagram forms a convex polyhedron in $\IIZ^3 \subset \IIZ^4$.
\footnote{See \cite{Aspinwall:1993nu,Martelli:2005tp,Martelli:2006yb} for more information on toric geometry}
The reduction from $\IIZ^4$ to $\IIZ^3$ is a consequence of the CY condition.
The crystal model follows from a T-duality of M-theory. We take the T-duality transformation along a $T^3\subset T^4$ 
aligned with the projection $\IIZ^4 \rightarrow  \IIZ^3$. 
This corresponds to the $x^{6,7,8}$ directions in Table 1.
By T-duality, we mean the element $t$ in the $SL(2,\IIZ)\times SL(3,\IIZ)$ duality 
group which acts as $t: \tau \equiv C_{(3)} + i \sqrt{g_{T^3}}
\rightarrow -1/\tau$.
The stack of $N$ M2-branes turns into a stack of $N$ M5-branes wrapping 
the dual $T^3$. 
We call them the $T$-branes. 
The degenerating circle fibers turn into another M5-brane 
extended along the (2+1)d world-volume and a non-trivial 3-manifold 
$S$ in $\IR^3 \times T^3$. We call it the $S$-brane. 
Preservation of supersymmetry requires that the $S$-brane
wrap a special Lagrangian submanifold of  
$\IR^3 \times T^3 = (\IIC^*)^3$, 
and that it is locally a plane in $\IR^3$ and a 1-cycle in $T^3$. 
The result is summarized in Table 1.
\begin{table}[htbp]
\label{brane3}
$$
\begin{array}{l|ccc|cccccc|cc}
\hline
           & 0 & 1 & 2 & 3 & 4 & 5 & 6 & 7 & 8 & 9 & 11\\
\hline
\mbox{M5} & \circ & \circ & \circ & & & & \circ & \circ & \circ \\
\mbox{M5} & \circ & \circ & \circ &  \multicolumn{6}{c|}{\Sigma} & & \\
\hline
\end{array}
$$
\caption{The brane configuration for M2 theories probing a toric CY$_4$.}
\end{table}

The crystal graph is  
the intersection locus between 
the $T$-branes and the $S$-brane projected onto the $T^3$. 
\fref{C4}, shows the crystal for $\IIC^4$. 
We have 4 bonds and 2 vertices. In crystal models each bond represents a ${\cal N}=2$
chiral field. As in dimer models, it is easy to read off the superpotential from
crystal models. Every vertex in the crystal
contributes a term in the superpotential, given by the product of all
the fields meeting at a vertex, with a positive sign for white
vertices and a negative sign for black ones.\footnote{As we explain below, current understanding of crystal models does not allow for the identification of gauge groups. Because of that, it is not clear how the gauge indices of chiral fields in superpotential terms are contracted.}
In Figure \ref{C4}, we see that we have four chiral fields and two
superpotential terms.
It is not clear how to read off the gauge group from the crystal model 
compatible with the CS theories proposed so far, 
though there has been partial success \cite{Kim:2007ic}. The proposal in \cite{Hanany:2008fj} seems to be promising for solving this problem. The ABJM model has four bifundamental chiral multiplets and the superpotential is identified with that
of the conifold (3+1)d theory \cite{ABJM}. This is in perfect agreement with the structure suggested by \fref{C4}. We will see later that there is another possibility for assigning gauge groups to the above crystal.

An important concept is that of a perfect matching. It is a collection
of bonds such that every node in the crystal belongs to exactly one bond.
In (3+1)d, it has been shown that there is a one to one correspondence between
perfect matchings in the dimer model and GLSM fields describing the moduli
space \cite{Franco:2006gc}. The same is true for the case of crystals, since it is straightforward
to show that perfect matchings are good variables for solving F-term equations.\footnote{Notice 
that this statement is not equivalent to saying that the correspondence between crystals and CY$_4$ singularities
is established. Although there is a natural proposal, there is no proof of how perfect matchings are positioned in a toric
diagram.} While all the calculations in the coming sections can be performed
{\it without} any reference to perfect matchings, it is sometimes practical to use this correspondence.

In addition, crystal models seem to be very useful in clarifying such
issues
as RG-flows, partial resolution and toric-duality in the (2+1)d setting. One can also use crystal models to work out the meson spectrum of the corresponding CFT$_3$, which is an important check of AdS$_4$/CFT$_3$ correspondence. In what follows, we use the information on the superpotential and
RG-flow obtained from crystal models to guide the construction of some (2+1)d theories.

\section{(2+1)d theories with and without (3+1)d parents} \label{4d}

\label{section_3d_not_4d}

\begin{figure}[h]
\begin{center}
\psfrag{C4 to conifold}[cc][][.8]{a) $\IIC^4 \to C(T^{11})$}
\psfrag{(C2/Z2)2 to F0}[cc][][.8]{b) $(\IIC^2/\IIZ_2)^2 \to C(F_0)$, $C(L^{222})$}
\psfrag{conifold x C to C2/Z2 x C}[cc][][.8]{c) $C(T^{11})\times \IIC \to \IIC^2/\IIZ_2 \times \IIC$}
\psfrag{M32 to dP0}[cc][][.8]{d) $C(M^{32}) \to C(dP_0)$}
\includegraphics[width=13cm]{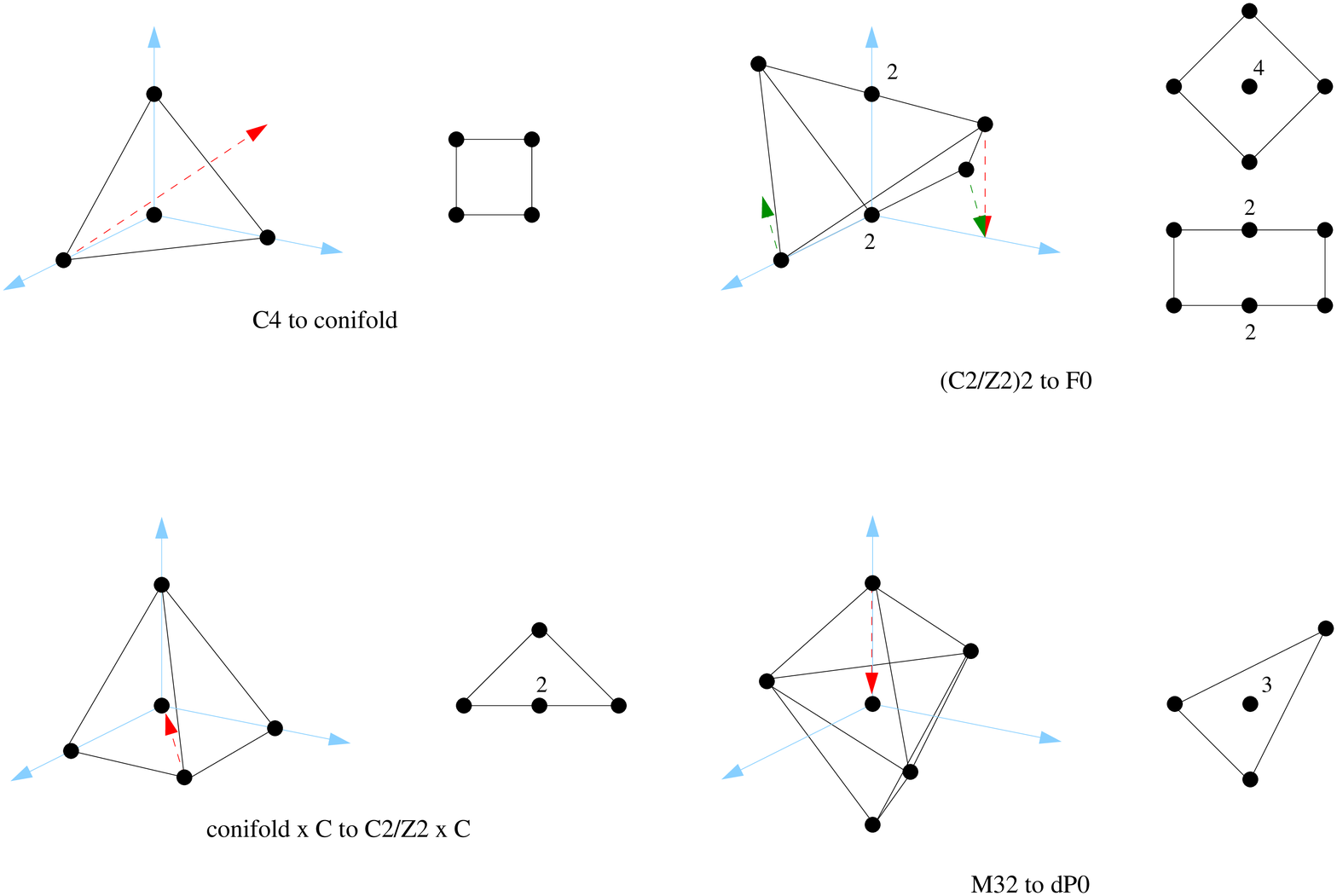}
\caption{3d toric diagrams and their projection to the 2d toric
  diagrams of the parents. 
(a) is the well known ABJM theory. It projects down to the conifold,
from which it borrows the quiver and 
superpotential. How the 2d toric diagram is `inflated' into a 3d one
depends on the choice of CS levels.
The number appearing in the toric diagram denotes the multiplicity of
the particular node. The multiplicity is one unless otherwise stated. (b) admits two projections (indicated in green and red) to toric diagrams coming from (3+1)d theories.}
\label{toric_projections}
\end{center}
\end{figure}

Recently, various authors discussed the possibility of generating (2+1)d
CS theories with toric 
CY$_4$ moduli spaces by taking theories with the same quiver diagrams and
superpotentials of 
theories in (3+1)d
\cite{Martelli,Ami,Ueda,Imamura2}. We
will refer 
to these models as theories with (3+1)d parents. While this represents an
interesting progress 
that allows the construction of an infinite number of new models, it
is not the generic situation and gives a reduced subset of theories for M2-branes over toric CY$_4$ manifolds.
It is possible to give a very intuitive characterization of all these
theories. 
They are theories whose 3d toric diagrams can be projected down to the
2d toric diagrams of 
the parent theories \cite{Hanany:2008fj}. All known theories
with (3+1)d parents satisfy this property.
When projected, an
important role is played by the multiplicity 
of GLSM fields \cite{Feng2,Feng1,Feng:2002zw}, namely the multiplicity of every node in the toric diagram has to match the one computed from a (3+1)d theory. 
It turns out that all (2+1)d CS theories with toric moduli spaces that
have been studied in the literature, 
even before the aforementioned references, fall into this subclass of
models with (3+1)d parents. 
\fref{toric_projections} shows a sample collection of those models and
their projections. 
Interestingly, some models like $(b)$ admit more than one
projection. If projected down, it gives the toric diagram of $C(F_0)$, a chiral $\IIZ_2$ orbifold of the conifold. If projected sideways, it gives the toric diagram of a non-chiral $\IIZ_2$ orbifold of the conifold (also denoted the cone over $L^{222}$). In these cases, 
the coincidence of moduli spaces can be verified by direct
computation. Interestingly, both theories have the same moduli space but, naively, different amounts of supersymmetry. While the first one seems to have ${\cal N}=2$, the second one has ${\cal N}=4$. It is natural to expect that SUSY is enhanced in the first model. We will explore these
issues in future work.

Conversely, the projection prescription gives us a way to identify
`pure' (2+1)d theories, 
namely those without (3+1)d parents. They are simply those whose toric
diagrams cannot be 
projected into 2d ones. A prototypical example is the cone over
$Q^{111}$. 
It is interesting to work out some pure (2+1)d theories in order to
understand their general structure 
and why they do not allow (3+1)d parents. This is the subject of our next
section.

\section{Gauge theory for $C(Q^{111})$}

\label{section_Q111}

\begin{figure}[h]
\begin{center}
\includegraphics[width=5cm]{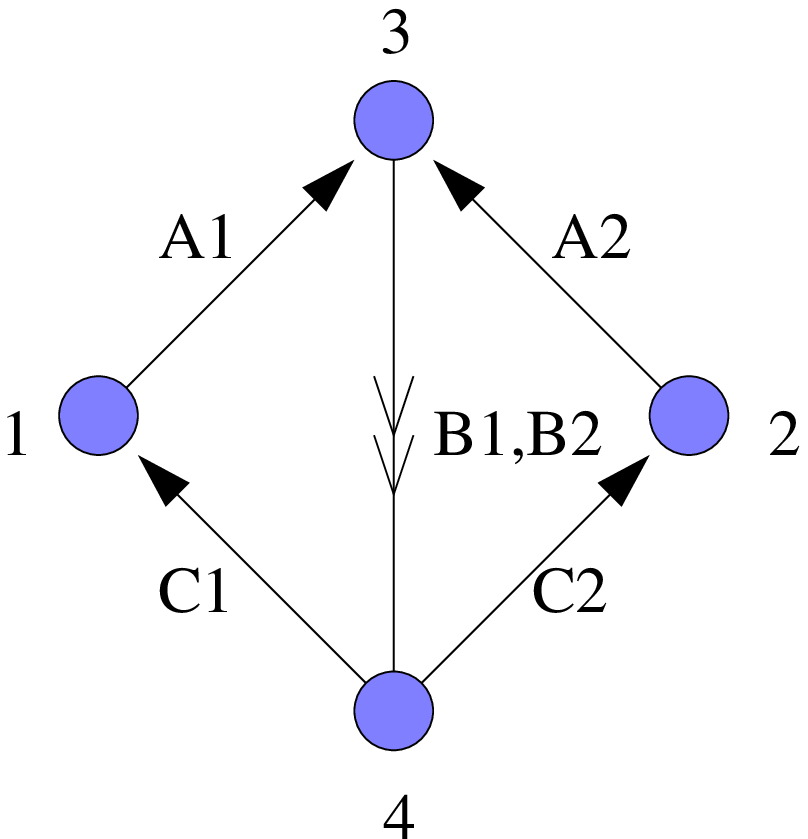}
\caption{Quiver diagram for $Q^{111}$.}
\label{quiver_Q111}
\end{center}
\end{figure}

We now construct the gauge theory for $C(Q^{111})$. Very much like the
conifold in (3+1)d, $C(Q^{111})$ is a great 
starting point due to its large symmetry. We will extract from
crystals as much information as possible, 
assuming their correctness. We will later see that they are indeed
right, by performing various checks, 
including the computation of the moduli space. From 
crystal model constructions \cite{Lee:2007kv,Kim:2007ic}, we know that $C(Q^{111})$ 
has:
\begin{itemize}
\item 6 chiral fields.
\item 2 non-vanishing superpotential terms of order 6.
\end{itemize}
Since the theory has two terms in the superpotential, there are
no restrictions on the abelian moduli space coming from $F$-terms. In other words, the superpotential vanishes in the abelian case and the master space is $\IIC^6$. Then, we must have 2 constraints from
D-terms, i.e., $G-2=2$ (with $G$ the number of gauge groups), thus we
also know the theory has 4 gauge groups. 
Finally, it is given by an $SU(2)^3/U(1)^2$ coset, which has $SU(2)^3 \times U(1)_R$ global symmetry. This structure appears clearly in our construction. The presence of 6 chiral fields is not surprising, since this is the minimal matter content we can think of in a theory with $SU(2)^3$ symmetry.
One can try to construct a theory that
meets all the requirements above. 
The constraints are so strong that the answer is basically
unique.\footnote{It is important to notice that there might exist dual descriptions of this theory, with different quivers, which share the same moduli space.} \fref{quiver_Q111} shows the proposed 
quiver diagram. We will then subject this theory to various tests.

The superpotential is given by
\beq
W= C_1 A_1 B_1 C_2 A_2 B_2 - C_1 A_1 B_2 C_2 A_2 B_1 ~.
\label{W_Q111}
\eeq
In this and coming expressions, color indices are contracted between adjacent fields and a trace is implicit. The theory has an explicit $SU(2)$ global symmetry under which $B_1$ and $B_2$
form a doublet, as well as a $U(1)_R$ symmetry. A useful intermediate step in the computation of the mesonic moduli space is the master space.\footnote{For short, we refer to the mesonic moduli space as just the moduli space in what follows.} To find it, we look for solutions of F-term equations without
imposing gauge invariance. As in any (toric) 
theory with two superpotential terms, we obtain $Q_F=0$ and the
GLSM fields (equivalently perfect matchings) are identified with the 
chiral fields. Let us call
$(p_1,p_2,p_3,p_4,p_5,p_6)=(A_1,A_2,B_1,B_2,C_1,C_2)$. The master
space is hence $\IIC^6$. 
We can construct the matrix of charges for GLSM fields (which in this case are equivalent to 
the chiral fields). The charges can be read from the quiver and are given by
\beq
\begin{array}{c|cccccc}
& \ \ A_1 \ \ & \ \ A_2 \ \ & \ \ B_1 \ \ & \ \ B_2 \ \ & \ \ C_1 \ \ & \ \ C_2 \ \ \\ \hline
\ \ Q_1 \ \ & 1 & 0 & 0 & 0 & -1 & 0 \\
\ \ Q_2 \ \ & 0 & 1 & 0 & 0 & 0 & -1 \\
\ \ Q_3 \ \ & -1 & -1 & 1 & 1 & 0 & 0 \\
\ \ Q_4 \ \ & 0 & 0 & -1 & -1 & 1 & 1 
\end{array}
\eeq

\begin{figure}[h]
\begin{center}
\includegraphics[width=4cm]{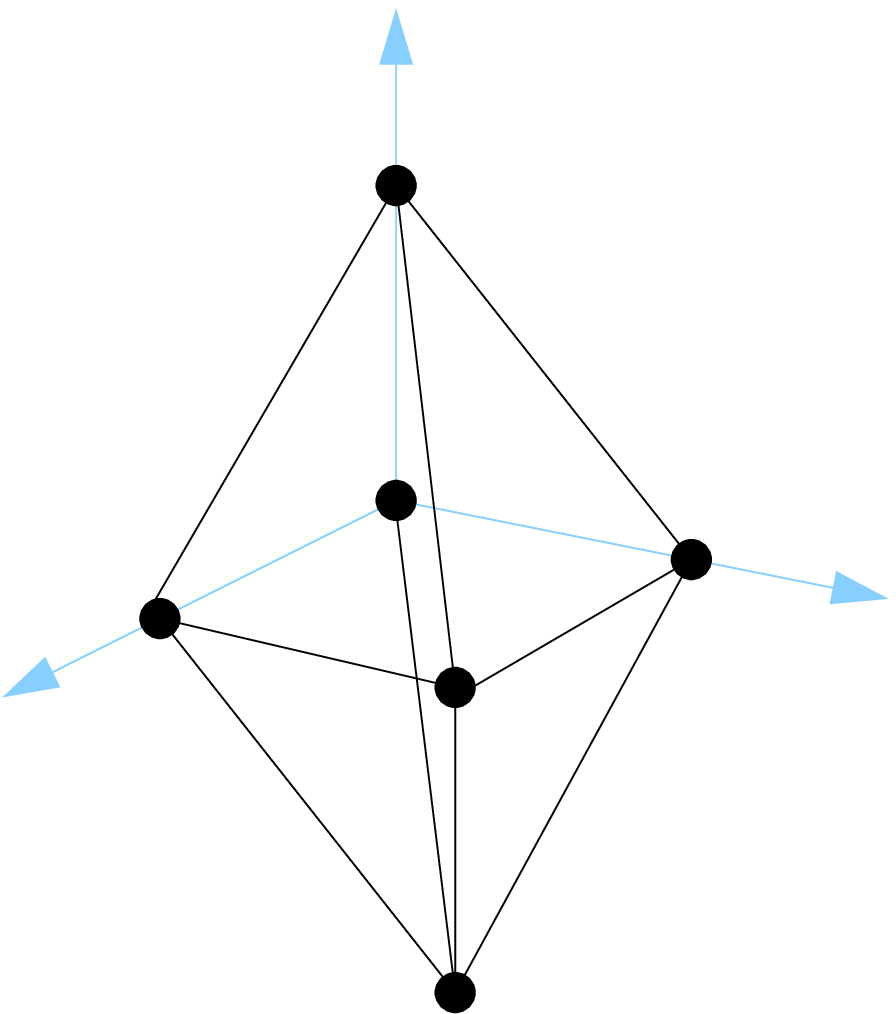}
\caption{Toric diagram for the $k=(1,1,-1,-1)$ theory.}
\label{toric_Q111}
\end{center}
\end{figure}

Different choices of the CS coefficients give interesting theories. We
are interested in breaking the symmetry 
of the master space down to global symmetry of $Q^{111}$, i.e. $SU(2)^3 \times U(1)$. There are only two choices that
lead to this symmetry at the level 
of the charge matrix. They are $(k,-k,0,0)$ and $(k,k,-k,-k)$. 
Since we do not want a further $\IIZ_k$ orbifold, we take $k=1$. 
Let us first consider $k=(1,-1,0,0)$. In this case, the two $U(1)$'s by which
we quotient can be taken to be 
$Q_3$ and $Q_4$. This gives rise to
\beq
Q_{D(1,-1,0,0)} = \left(\begin{array}{cccccc} 
-1 & -1 & 1 & 1 & 0 & 0 \\
\ \ 0 \ \ & \ \ 0 \ \ & \ \ -1 \ \ & \ \ -1 \ \ & \ \ 1 \ \ & \ \ 1 \ \ 
\end{array}\right)
\eeq
We clearly see that this charge matrix breaks the global symmetry of the master space from $U(6)$ down to $SU(2)^3 \times U(1)$, as desired. The pairs $(A_1,A_2)$, $(B_1,B_2)$ and $(C_1,C_2)$ transform as doublets of each of the $SU(2)$ factors.
The toric diagram is given by the kernel of this matrix and is equal to
\beq
G^T_{(1,-1,0,0)}=\left(\begin{array}{cccccc} 
\ \ 1 \ \ & \ \ 0 \ \ & \ \ 1 \ \ & \ \ 0 \ \ & \ \ 0 \ \ & \ \ 1 \ \ \\ 
1 & 0 & 1 & 0 & 1 & 0 \\
0 & 0 & -1 & 1 & 0 & 0 \\
-1 & 1 & 0 & 0 & 0 & 0
\end{array}\right)
\eeq
All columns add up to 1, as usual. We can drop, for example, the
fourth row and plot the toric diagram. 
The result is shown \fref{toric_Q111} and is precisely the one for $C(Q^{111})$. 

We now repeat the analysis for $k=(1,1,-1,-1)$. In this case, we quotient by $Q_1+Q_3$ and $Q_1+Q_4$, which gives 

\beq
Q_{D(1,1,-1,-1)} = \left(\begin{array}{cccccc} 
0 & -1 & 1 & 1 & -1 & 0 \\
\ \ 1 \ \ & \ \ 0 \ \ & \ \ -1 \ \ & \ \ -1 \ \ & \ \ 0 \ \ & \ \ 1 \ \ 
\end{array}\right)
\eeq
Once again, the $SU(2)^3 \times U(1)$ symmetry is clear from this matrix. The doublets are now different from the previous case, and are given by $(A_1,C_2)$, $(A_2,C_1)$ and $(B_1,B_2)$. Taking the kernel we obtain

\beq
G^T_{(1,1,-1,-1)}=\left(\begin{array}{cccccc} 
\ \ -1 \ \ & \ \ 0 \ \ & \ \ 0 \ \ & \ \ 0 \ \ & \ \ 0 \ \ & \ \ 1 \ \ \\ 
0 & -1 & 0 & 0 & 1 & 0 \\
1 & 1 & 0 & 1 & 0 & 0 \\
1 & 1 & 0 & 0 & 0 & 0
\end{array}\right)
\eeq
It is straightforward to see that it also corresponds to the toric diagram in \fref{toric_Q111}.

Let us provide some argument of why this theory does not come from
a (3+1)d parent. 
From \fref{quiver_Q111}, we see that the quiver contains nodes with a single incoming and a single outgoing arrow.
Such nodes correspond to $N_f=N_c$ gauge groups and possibly 
generate dynamical scales, not leading to a CFT.

\section{A Klebanov-Witten RG flow}

\label{section_dP3}

\begin{figure}[h]
\begin{center}
\includegraphics[width=5cm]{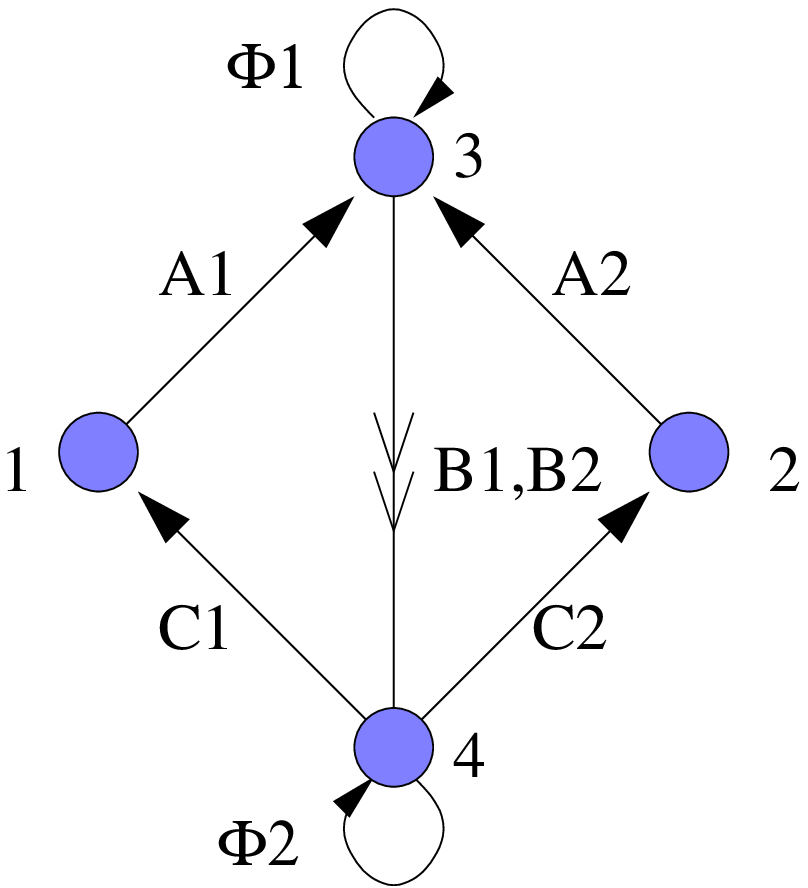}
\caption{Quiver diagram for $C(dP_3)\times \IIC$.}
\label{quiver_dP3}
\end{center}
\end{figure}

Using crystals, the authors of \cite{Kim:2007ic} have proposed some
Klebanov-Witten type RG flows 
\cite{Klebanov} connecting theories, which result from adding
adjoint masses. 
The adjoint masses come from twisting bonds in the crystal. In
particular, it is  suggested there 
should exist such a flow between $C(dP_3) \times \IIC$ and
$C(Q^{111})$. We now investigate this flow and 
use it to determine the gauge theory for $C(dP_3) \times \IIC$. We go a
step beyond \cite{Kim:2007ic} and 
propose the quiver for this model, which is shown in
\eref{quiver_dP3}. It is obtained by undoing the 
RG flow that we now explain.

\begin{figure}[h]
\begin{center}
\includegraphics[width=4cm]{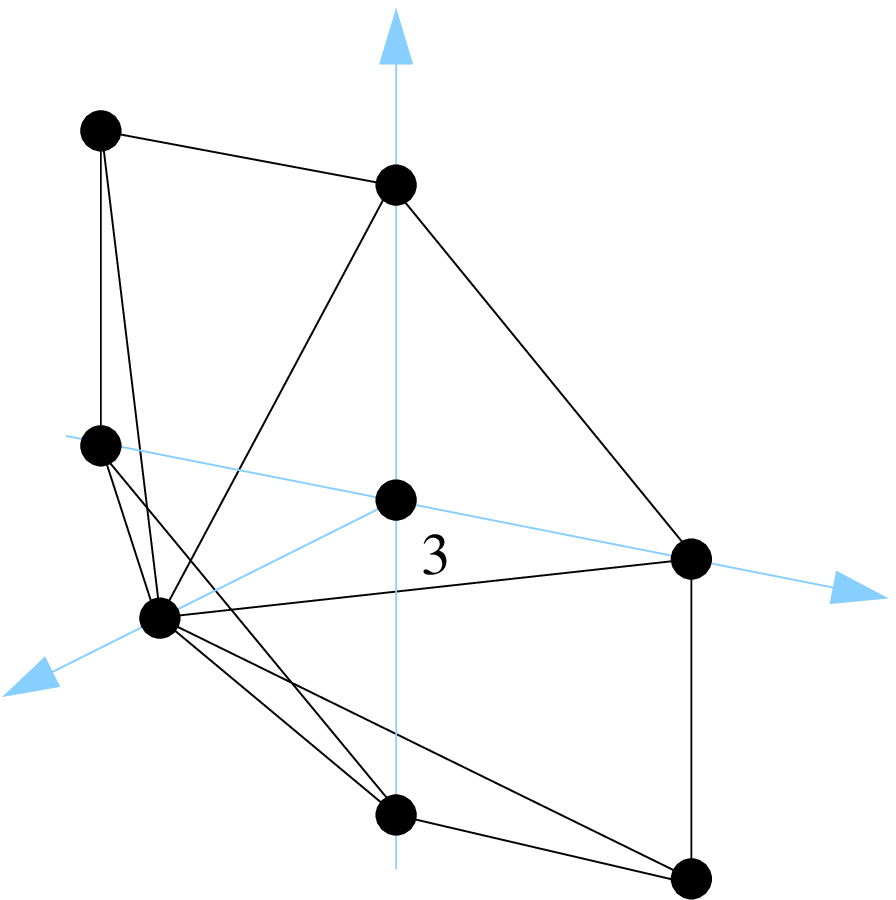}
\caption{Toric diagram for the $k=(1,1,-1,-1)$ theory.}
\label{toric_dP3}
\end{center}
\end{figure}

The superpotential is \footnote{Contrary to \cite{Kim:2007ic} where the
 superpotential is known in the abelian limit, we know the gauge indices of
 all fields. 
We sort the fields in the superpotential accordingly.}
\beq
W=\phi_1 (B_1 C_1 A_1 - B_2 C_2 A_2) - \phi_2 (C_1 A_1 B_1 - C_2 A_2 B_2) ~.
\label{W_dP3}
\eeq
The RG flow is triggered by the following mass term
\beq
\Delta W = -(\phi_1^2 - \phi_2^2) .
\eeq
It is straightforward to verify that, by integrating out the massive
fields $\phi_1$ and $\phi_2$, 
we recover \eref{W_Q111} up to an unimportant overall multiplicative constant. 
This is indeed very encouraging. Let us now check that the theory with
quiver diagram in \fref{quiver_dP3} 
and superpotential \eref{W_dP3} has
$C(dP_3)\times \IIC$ as its moduli space for some choice of CS levels.
As before, it is a straightforward exercise to write down the matrix
translating quiver fields to GLSM fields. 
It is given by
\beq
P=\left(\begin{array}{c|cccccccccc}
\ \ \ \ \ & \ \ p_1 \ \ & \ \ p_2 \ \ & \ \ p_3 \ \ & \ \ p_4 \ \ & \
\ p_5 \ \ & \ \ p_6 \ \ & \ \ 
p_7 \ \ & \ \ p_8 \ \ & \ \ p_9 \ \ & \ \ p_{10} \ \ \\
\hline
A_1 & 1 & 1 & 1 & 0 & 0 & 0 & 0 & 0 & 0 & 0 \\
A_2 & 1 & 0 & 0 & 1 & 0 & 0 & 1 & 0 & 0 & 0 \\
B_1 & 0 & 0 & 0 & 1 & 1 & 1 & 0 & 0 & 0 & 0 \\
B_2 & 0 & 1 & 0 & 0 & 1 & 0 & 0 & 1 & 0 & 0 \\
C_1 & 0 & 0 & 0 & 0 & 0 & 0 & 1 & 1 & 1 & 0 \\
C_2 & 0 & 0 & 1 & 0 & 0 & 1 & 0 & 0 & 1 & 0 \\
\phi_1 & 0 & 0 & 0 & 0 & 0 & 0 & 0 & 0 & 0 & 1 \\
\phi_2 & 0 & 0 & 0 & 0 & 0 & 0 & 0 & 0 & 0 & 1 
\end{array}\right)
\label{P_dP3}
\eeq
This determines the charge matrix $Q_F = Ker(P)$ encoding the F-term equations.
\beq
Q_F=\left(\begin{array}{cccccccccc}
 \ \ p_1 \ \ & \ \ p_2 \ \ & \ \ p_3 \ \ & \ \ p_4 \ \ & \ \ p_5 \ \ &
 \ \ p_6 \ \ & \ \ p_7 \ \ & 
\ \ p_8 \ \ & \ \ p_9 \ \ & \ \ p_{10} \ \ \\ \hline
1 & 0 & -1 & 0 & 0 & 0 & -1 & 0 & 1 & 0 \\
1 & -1 & 0 & 0 & 0 & 0 & -1 & 1 & 0 & 0 \\
1 & 0 & -1 & -1 & 0 & 1 & 0 & 0 & 0 & 0 \\
1 & -1 & 0 & -1 & 1 & 0 & 0 & 0 & 0 & 0 
\end{array} \right)
\label{QF_dP3} 
\eeq
From \eref{P_dP3}, we can determine how GLSM fields are charged under
the four quiver $U(1)$'s. 
This is given by
\beq
\begin{array}{c|cccccccccc}
\ \ \ \ \ & \ \ p_1 \ \ & \ \ p_2 \ \ & \ \ p_3 \ \ & \ \ p_4 \ \ & \
\ p_5 \ \ & \ \ p_6 \ \ & \ \ p_7 \ \ 
& \ \ p_8 \ \ & \ \ p_9 \ \ & \ \ p_{10} \ \ \\ \hline
Q_1 & 1 & 0 & 0 & 0 & 0 & 0 & -1 & 0 & 0 & 0 \\
Q_2 & 0 & 0 & 0 & 0 & 0 & 0 & 1 & 0 & -1 & 0 \\
Q_3 & -1 & 0 & 0 & 0 & 1 & 0 & 0 & 0 & 0 & 0 \\
Q_4 & 0 & 0 & 0 & 0 & -1 & 0 & 0 & 0 & 1 & 0 
\end{array}
\eeq
Since the CS levels are not affected by the RG flow, once again we are
interested in looking at the theory with 
$k=(1,1,-1,-1)$. This tells us that we can impose the D-terms for
$Q_1+Q_3$ and $Q_1+Q_4$. 
\beq
Q_D=\left(\begin{array}{cccccccccc}
 \ \ p_1 \ \ & \ \ p_2 \ \ & \ \ p_3 \ \ & \ \ p_4 \ \ & \ \ p_5 \ \ &
 \ \ p_6 \ \ & \ \ p_7 \ \ & 
\ \ p_8 \ \ & \ \ p_9 \ \ & \ \ p_{10} \ \ \\ \hline
0 & 0 & 0 & 0 & 1 & 0 & -1 & 0 & 0 & 0 \\
1 & 0 & 0 & 0 & -1 & 0 & -1 & 0 & 1 & 0 
\end{array}\right)
\label{QD_dP3}
\eeq
The total charge matrix is obtained from concatenating \eref{QF_dP3}
and \eref{QD_dP3}. 
The toric diagram is again given by
\beq
G^T = Ker (Q_{tot})= 
\left(\begin{array}{cccccccccc}
 \ \ p_1 \ \ & \ \ p_2 \ \ & \ \ p_3 \ \ & \ \ p_4 \ \ & \ \ p_5 \ \ &
 \ \ p_6 \ \ & \ \ p_7 \ \ 
& \ \ p_8 \ \ & \ \ p_9 \ \ & \ \ p_{10} \ \ \\ \hline
0 & 0 & 0 & 0 & 0 & 0 & 0 & 0 & 0 & 1 \\
-1 & -1 & 0 & 0 & 0 & 1 & 0 & 0 & 1 & 0 \\
0 & 1 & 0 & -1 & 0 & -1 & 0 & 1 & 0 & 0 \\
2 & 1 & 1 & 2 & 1 & 1 & 1 & 0 & 0 & 0 
\end{array} \right)
\eeq
The result is represented in \fref{toric_dP3}, where we use the
first three rows of the previous matrix. 
This is indeed the toric diagram for $C(dP_3) \times \IIC$.

The multiplicity of GLSM fields in \fref{toric_dP3} does not project
down to any of the toric diagrams 
that arise from $dP_3$ quivers \cite{Feng:2002zw}. We also know this is the
case because, otherwise, 
the gauge theory would have six gauge groups and a completely
different quiver. Like $C(Q^{111})$ this is 
a pure (2+1)d theory.

\section{More examples}

\subsection{Another pair of theories connected by an RG flow}

\label{section_other_pair}

We now present a similar pair of theories connected by an RG flow,
also anticipated in \cite{Lee:2007kv}. 
The two theories correspond to $D_3$ and $\IIC^3/(\IIZ_2 \times \IIZ_2) \times \IIC$. 
Crystal model suggest that the superpotential of the theories
corresponding to $\IIC^3/(\IIZ_2 \times \IIZ_2) \times \IIC$ 
are the same as those of $C(dP_3) \times \IIC$ in the abelian limit (namely when fields are no longer matrices and ordering becomes unimportant). I.e., nodes in both crystals combine the same fields. This hints that the matter contents are related by suitable flips of the charges. We expect $D_3$ and $C(Q^{111})$ to be connected in a similar way.

\begin{figure}[h]
\begin{center}
\includegraphics[width=5cm]{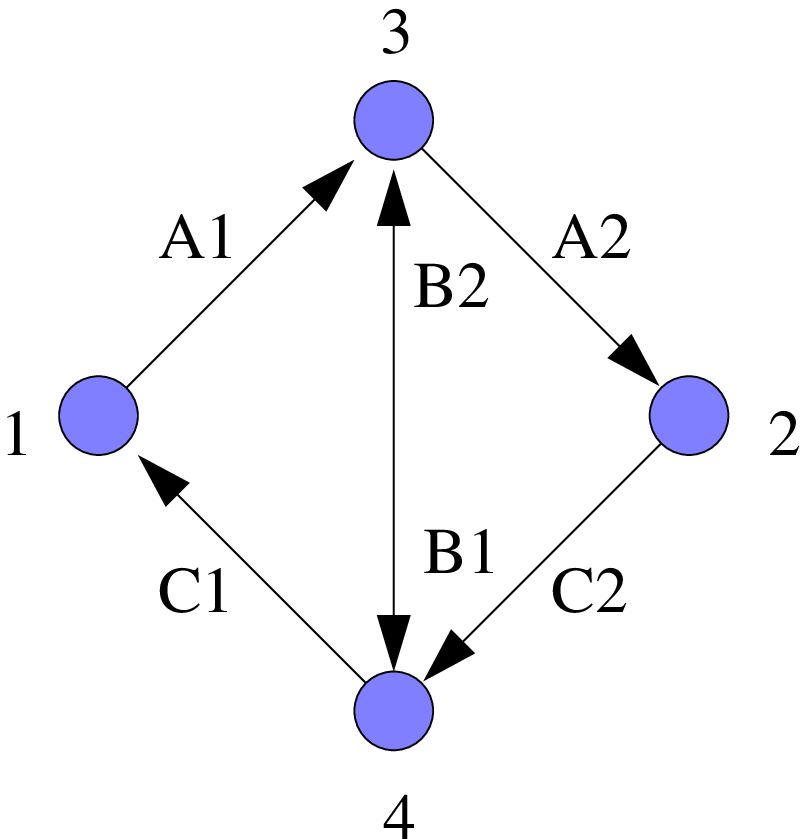}
\caption{Quiver diagram for $D_3$.}
\label{quiver_D3}
\end{center}
\end{figure}

Let us first consider $D_3$. Its quiver is shown in
\fref{quiver_D3}. It is obtained from the 
$C(Q^{111})$ quiver by flipping half of the arrows. The superpotential is
\beq
W= C_1 A_1 B_1 B_2 A_2 C_2 - B_1 C_1 A_1 A_2 C_2 B_2 ~.
\label{W_D3}
\eeq
This superpotential follows form crystal models and, as we explained, can be obtained from the superpotential of $C(Q^{111})$ by changing the
order of fields according to the charge assignments.

\begin{figure}[h]
\begin{center}
\includegraphics[width=4cm]{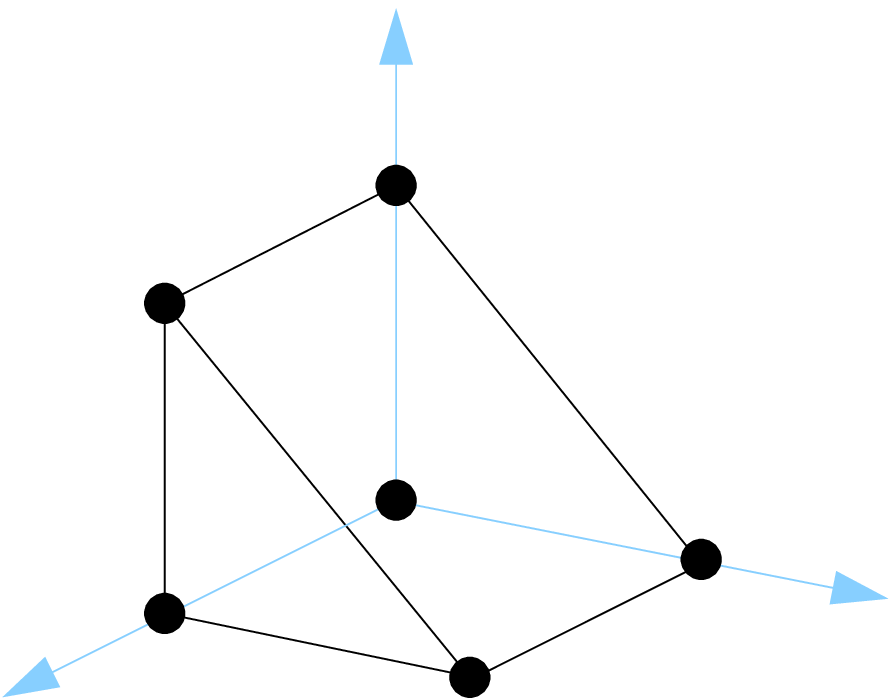}
\caption{Toric diagram for the $k=(1,1,-1,-1)$ theory.}
\label{toric_D3}
\end{center}
\end{figure}

Since we have only two superpotential terms, $Q_F=0$, as for
$Q^{111}$, and GLSM fields are identified with chiral fields. 
The quiver $U(1)$ charges are given by
\beq
\begin{array}{r|cccccccc}
  & A_1 & C_1 & A_2 & C_2 & B_1 & B_2 \\ \hline
Q_1 & 1 & -1 & 0 & 0 & 0 & 0 \\
Q_2 & 0 & 0 & -1 & 1 & 0 & 0 \\
Q_3 & -1 & 0 & 1 & 0 & 1 & -1 \\
Q_4 & 0 & 1 & 0  & -1 & -1 & 1 
\end{array}
\eeq
As for $Q^{111}$, there are two choices of CS levels that produce the desired moduli space: $k=(1,1,-1,-1)$ and $k=(1,-1,0,0)$. We analyze $k=(1,1,-1,-1)$, the other option is analogous. We quotient by $Q_1+Q_3$ and $Q_1+Q_4$, given by the matrix
\beq
Q_D = \left(\begin{array}{cccccc} 
0 & -1 & 1 & 0 & 1 & -1 \\ 
\ \ 1 \ \ & \ \ 0 \ \ & \ \ 0 \ \ & \ \ -1 \ \ & \ \ -1 \ \ & \ \ 1 \ \ \\
\end{array}\right)
\eeq
In this case, $Q_{tot}=Q_D$. Its kernel determines the toric diagram matrix
\beq
G^T=\left(\begin{array}{cccccc} 
\ \ -1 \ \ & \ \ -1 \ \ & \ \ 0 \ \ & \ \ 0 \ \ & \ \ 0 \ \ & \ \ 1 \ \ \\ 
1 & 1 & 0 & 0 & 1 & 0 \\
1 & 0 & 0 & 1 & 0 & 0 \\
0 & 1 & 1 & 0 & 0 & 0
\end{array}\right)
\eeq
This matrix corresponds to the toric diagram for $D_3$. \fref{toric_D3}
plots its last three rows.

In passing, we note that it is easy to identify another theory whose moduli space is $D_3$. It arises from the (3+1)d parent theory of the cone over the Suspended Pinch Point(SPP), whose toric diagram can be obtained from that of $D_3$ by a suitable projection. In this case, the gauge theory has only three gauge groups \cite{Morrison:1998cs} and the CS couplings are $k=(1,-1,0)$, with zero in one of the gauge groups without the adjoint \cite{Hanany:2008fj}.
This example shows a behavior that we expect to be generic, the same moduli space arises from theories with and without (3+1)d parents. Furthermore, these theories can have a different number of gauge groups. 

\begin{figure}[h]
\begin{center}
\includegraphics[width=5cm]{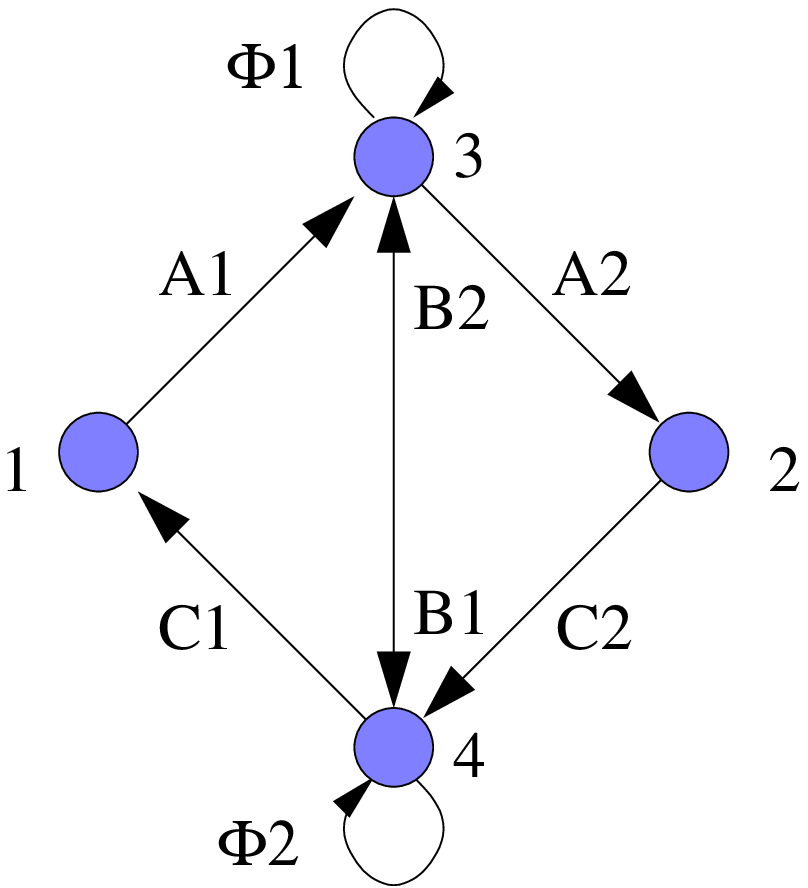}
\caption{Quiver diagram for $\IIC^3/(\IIZ_2 \times \IIZ_2)\times \IIC$.}
\label{quiver_Z2_Z2}
\end{center}
\end{figure}

We now propose a theory for $\IIC^3/(\IIZ_2 \times \IIZ_2) \times
\IIC$. We obtain its quiver from the one of $C(dP_3) \times \IIC$, shown in
\fref{quiver_dP3}, 
by flipping the direction of $A_2$, $B_2$ and $C_2$. The quiver diagram is shown in \fref{quiver_Z2_Z2}.

\begin{figure}[h]
\begin{center}
\includegraphics[width=5cm]{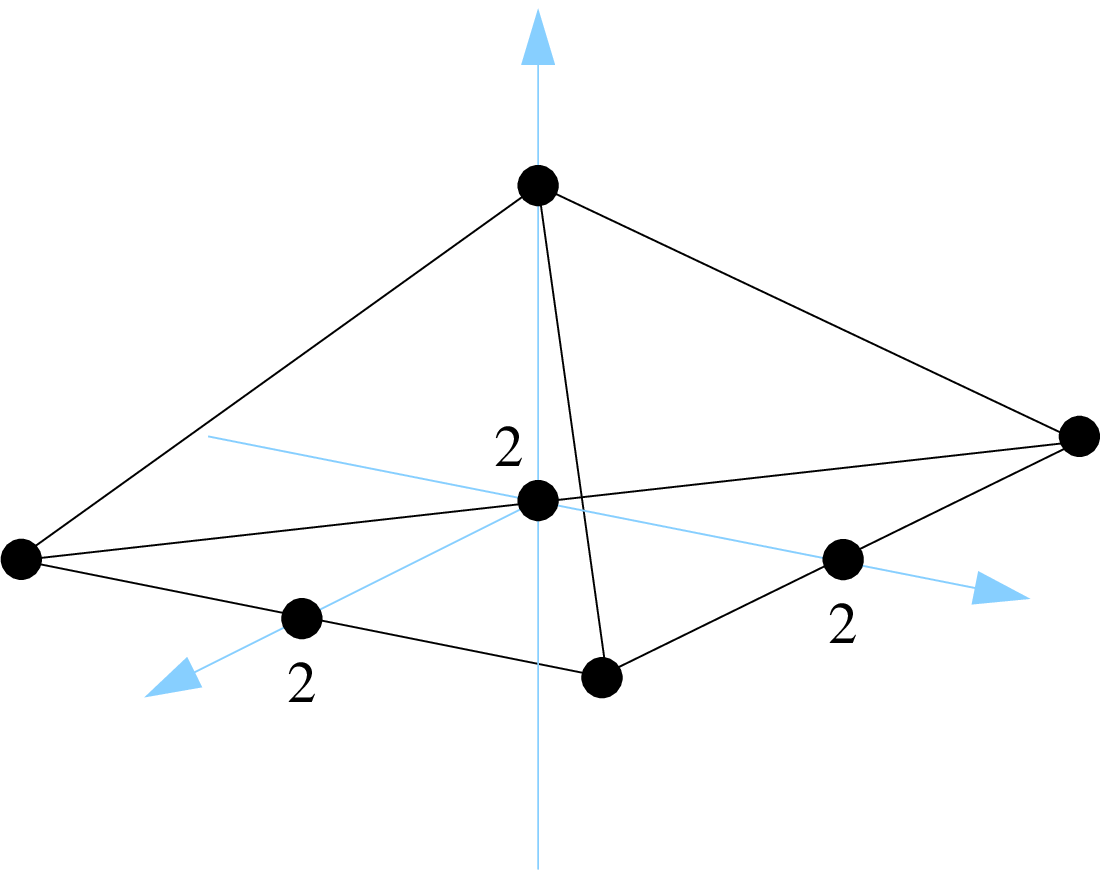}
\caption{Toric diagram for the $k=(1,1,-1,-1)$ theory.}
\label{toric_Z2_Z2}
\end{center}
\end{figure}

The superpotential is
\begin{equation}
W= \phi_1 (B_1 C_1 A_1-A_2 C_2 B_2) +\phi_2 (B_2 A_2 C_2 -C_1 A_1
B_1)
\label{W_Z2_Z2}
\end{equation}
As explained before, the superpotential of $\IIC^3/(\IIZ_2 \times
\IIZ_2)\times \IIC$ is the same as $C(dP_3) \times \IIC$ in the abelian
case. Because of this, the $P$ and $Q_F$ matrices as are the same of  $C(dP_3) \times \IIC$,
\eref{P_dP3} and \eref{QF_dP3}. 
The quiver $U(1)$ charges correspond to
\beq
\begin{array}{r|cccccccccccc}
  & p_1 & p_2 & p_3 & p_4 & p_5 & p_6 & p_7 & p_8 &
  p_9 & p_{10} \\ \hline
Q_1 & 0 & 0 & 1 & 1 & 0 & -1 & -1 & 0 & 0 & 0 \\
Q_2 & 0 & -1 & 1 & 0 & 0 & 0 & -1 & 1 & 0 & 0 \\
Q_3 & 0 & -1 & 0  & 1 & 0 & 0 & 0 & 0 & 0 & 0 \\
Q_4 & 0 & 0 & 0 & 0 & 0 & -1 & 0 &  1 & 0 & 0
\end{array}
\eeq
Let us consider $k=(1,1,-1,-1)$ (once again, $k=(1,-1,0,0)$ 
gives the same moduli space). We then consider $Q_1+Q_3$ and $Q_1+Q_4$, which give
\begin{equation}
Q_D= \left ( \begin{tabular}{cccccccccc}
$p_1$ & $p_2$ & $p_3$ & $p_4$ & $p_5$ & $p_6$ & $p_7$ & $p_8$ &
  $p_9$ & $p_{10} $ \cr \hline
 0 & -1 & 1 & 2 &  0 & -1 & -1 & 0 & 0 & 0 \cr
 0 & 0 & 1  & 1 & 0 & -2 & -1 & 1 & 0 & 0 
\end{tabular}  \right )
\end{equation}
Combining $Q_F$ and $Q_D$ and finding its kernel, we get 
\begin{equation}
G^T=Ker(Q_{tot})= \left ( \begin{tabular}{cccccccccc}
$p_1$ & $p_2$ & $p_3$ & $p_4$ & $p_5$ & $p_6$ & $p_7$ & $p_8$ &
  $p_9$ & $p_{10} $ \cr \hline
 1 & 0 & 1 & 0 & -1 & 0 & 1 & 0 & 1 & 0 \cr
 0 & 0 & -1 & 1 & 1 & 0 & 1 & 1 & 0 & 0  \cr
 0 & 1 & 1 & 0 & 1 & 1 & -1 & 0 & 0 & 0  \cr
 0 & 0 & 0 & 0 & 0 & 0 & 0 & 0 & 0 & 1
\end{tabular}  \right ).
\end{equation}
Removing, for example, the first row gives the toric diagram of
$\IIC^3/(\IIZ_2 \times \IIZ_2) \times \IIC$ as shown in \fref{toric_Z2_Z2}.

There is a further check we can perform on the two theories we have just introduced. In \cite{Kim:2007ic}, it is suggested that there exists an
Klebanov-Witten type RG-flow connecting 
$\IIC^3/(\IIZ_2 \times \IIZ_2)\times \IIC$ and $D_3$ theories. Indeed, it
is easy to check that adding 
$(\phi_2^2-\phi_1^2)$ to \eref{W_Z2_Z2} and integrating out the
massive fields, 
we obtain \eref{W_D3}.

\subsection{$\IIC^3/(\IIZ_N \times \IIZ_N) \times \IIC$ orbifold}

\label{section_ZN_ZN}

We can extend our results for $\IIC^3/(\IIZ_2 \times \IIZ_2) \times \IIC$
and give a proposal for a general {$\IIC^3/(\IIZ_N \times \IIZ_N) \times
  \IIC$ orbifold. 
The theory contains $2N$ gauge groups and $4N$ matter fields given by
\beq
\begin{array}{ccc}
X_i & \ \ \ \ \ \ & (\fund_{2(i-1)},\antifund_{2i-1}) \\
Y_i & \ \ \ \ \ \ & (\fund_{2i-1},\antifund_{2i}) \\
Z_i & \ \ \ \ \ \ & (\fund_{2i},\antifund_{2(i-1)}) \\
\phi_i & \ \ \ \ \ \ & \mbox{Adj}_{2i}
\end{array}
\eeq
with $i=1,\ldots,N$ and nodes in the quiver identified by $\mod(2N)$. The superpotential is
\beq
W=\sum_{i=1}^N \phi_i (X_{i+1}Y_{i+1}Z_{i+1}-Z_i X_i Y_i) ~.
\eeq
From this superpotential, we can use Kasteleyn matrix techniques to
determine that 
the number of GLSM fields is $3^N + 1$.\footnote{It is interesting to compare this number with 
the $2^N+1$ GLSM fields of $\IIC^2/\IIZ_N \times \IIC$ orbifolds
\cite{Feng:2002zw}.} 
Because of this, it is computationally difficult to verify this
proposal for large $N$. 
In Appendix \ref{A}, we confirm it explicitly for $N=3$. The notation
in the previous section for $\IIC^3/(\IIZ_2 \times \IIZ_2) \times \IIC$ translates into the general notation as follows: $(A_1,A_2,B_1,B_2,C_1,C_2)\to (Y_1,X_1,Z_1,Z_2,Y_2,X_2)$ and nodes are relabeled according to $(1,2,3,4) \to (1,3,2,4)$.

It is interesting to consider the $N=1$ case, since it provides an alternative to ABJM for M2-branes on $\IIC^4$. The model has a $U(N_1) \times U(N_2)$ gauge group with $X$ transforming
as $(N_2,\bar{N}_1)$, $Y$ as $(N_1,\bar{N}_2)$ and two adjoints
$\phi_1$ and $Z=\phi_2$ of $U(N_2)$. The superpotential is given by 
\beq
W= \phi_1 \, X \, Y \, \phi_2-\phi_2 \, X \, Y \, \phi_1 ~.
\eeq
The  moduli space is $\IIC^4$ for CS levels $(1,-1)$. See \cite{Hanany:2008fj} for the same theory, but derived by other methods. It would be interesting to understand how supersymmetry is enhanced in this model.

\section{Partial resolution}

\label{section_partial_resolution}

\subsection{Partial resolution in CS theories}

Different geometries and their dual gauge theories can be connected by partial resolution. Partial resolution works in this case very similarly to (3+1)d, with a few new features that we now discuss.

We can turn on FI parameters for any of the gauge groups, with the consequent modification of the D-term equations. The $G-2$ ones that originally vanish are of most importance. As a result of the FI terms, some chiral fields in the quiver (equivalently the corresponding GLSM fields) acquire vevs. These vevs higgs the theory at low energies and can also give mass to some of the chiral fields, which have to be integrated out.

It is interesting to notice that for the specific case of manifolds of the form CY$_3 \times \IIC$, the number of 
possible partial resolutions is smaller than for CY$_3$. The reason for this is twofold. The
CY$_3 \times \IIC$ theory has generally less gauge groups than the CY$_3$ counterpart\footnote{A simple example that falls into this category but does not satisfy this rule is $\IIC^4 =\IIC^3 \times \IIC$. The ABJM theory (the theory for $\IIC^4$) has one gauge group more than 
${\mathcal N}=4$ SYM in (3+1)d (the theory for $\IIC^3$).} and only $G-2$ independent FI terms result in resolutions.

We also need to take care of the CS couplings. As we now show, whenever two gauge groups are higgsed to the diagonal subgroup by a bifundamental vev, the resulting CS coupling is the sum of the original ones. Suppose some field with charges $(-1,1)$ under gauge groups ${\cal A}^1$ and ${\cal A}^2$, whose CS couplings are $k_1$ and $k_2$, acquires a vev. For its scalar component, the covariant derivative is

\begin{equation}
D_{\mu}\Phi=\partial_{\mu}\Phi-i({\cal A}^1_{\mu}-{\cal A}^2_{\mu})\Phi\ ~.
\end{equation}
The combination ${\cal A}^-_{\mu}={\cal A}^1_{\mu}-{\cal A}^2_{\mu}$ becomes massive. We call $m$ its mass. The relevant piece of the action is

\begin{equation}
S=\int d^3x\, k_1 \epsilon^{\mu\nu\rho} {\cal A}^1_{\mu}\partial_{\nu}{\cal A}^1_{\rho}+k_2\epsilon^{\mu\nu\rho}{\cal A}^2_{\mu}\partial_{\nu}{\cal A}^2_{\rho} - m^2 ({\cal A}_{\mu}^1-{\cal A}_{\mu}^2)^2+\cdots\ .
\end{equation}
Defining ${\cal A}^{\pm}={\cal A}^1\pm {\cal A}^2$ and $k_{\pm}=k_1 \pm k_2$, we get  

\begin{equation}
S=\int d^3x\, k_+\epsilon^{\mu\nu\rho} {\cal A}^+_{\mu}\partial_{\nu}{\cal A}^+_{\rho}+k_+\epsilon^{\mu\nu\rho}{\cal A}^-_{\mu}\partial_{\nu}{\cal A}^-_{\rho} +2k_-\epsilon^{\mu\nu\rho} {\cal A}^-_{\mu}\partial_{\nu}{\cal A}^+_{\rho}-m^2({\cal A}^-_{\mu})^2+\cdots\ .
\end{equation}

At energies well below $m$, we can proceed to integrate out ${\cal A}^-$. The equation of motion reads 

\begin{equation}
k_-\epsilon^{\mu\nu\rho}\partial_{\nu}{\cal A}^+_{\rho}+k_+\epsilon^{\mu\nu\rho}\partial_{\nu}{\cal A}^-_{\rho}=m^2{\cal A}_{\mu}^- ~.
\end{equation}
At energies well below $m$, we can consider ${\cal A}^-$ is constant. Then, the previous expression reduces to
\begin{equation}
k_-\epsilon^{\mu\nu\rho}\partial_{\nu}{\cal A}^+_{\rho}\sim m^2{\cal A}_{\mu}^-
\end{equation}
and
\begin{equation}
S\sim\int d^3x\, k_+\epsilon^{\mu\nu\rho} {\cal A}^+_{\mu}\partial_{\nu}{\cal A}^+_{\rho}+2k_-\epsilon^{\mu\nu\rho} {\cal A}^-_{\mu}\partial_{\nu}{\cal A}^+_{\rho}-m^2({\cal A}^-_{\mu})^2+\cdots\ .
\end{equation}
Plugging the approximate solution to the equation of motion we get
\begin{equation}
S=\int d^3x\, k_+\epsilon^{\mu\nu\rho} {\cal A}^+_{\mu}\partial_{\nu}{\cal A}^+_{\rho}-\frac{k_-^2}{2m^2}F_{\mu\nu}^+F_{\mu\nu}^+\cdots\ .
\end{equation}
As anticipated, we get a CS coupling for the surviving gauge field whose CS level is the sum of the Higgsed CS levels. In addition, there is a Maxwell term that vanishes in the IR limit (equivalently in the $m\to \infty$ limit).

\subsection{Connections between models}

We now investigate the web of connections that result from partial resolutions between the theories we have studied. With this goal in mind, the list of partial resolutions 
we considered is certainly not exhaustive.

By now, we expect the reader to be familiar with the kind of matrices that arise when analyzing these models
from a toric geometry perspective. Hence, for the brevity of the presentation, we just state the quiver vevs
that are turned on (working out the corresponding vevs for GLSM fields is straightforward) and the results. 

$\IIC^3/(\IIZ_3 \times \IIZ_3) \times \IIC$ is resolved down to $\IIC^3/(\IIZ_2 \times \IIZ_2) \times \IIC$ by turning
on vevs for $X_1$ and $Z_1$. The CS levels match the ones we have studied. $\IIC^3/(\IIZ_2 \times \IIZ_2) \times \IIC$ can be resolved to $C(T^{11}) \times \IIC$ by vevs of $A_1$ and $A_2$. The quiver diagram is shown in \fref{conifold}.a, $k=(1,-1)$ and the superpotential is

\begin{equation}
W= \phi_1 (B_1 C_1 -C_2 B_2) +\phi_2 (B_2 C_2 -C_1 B_1) ~.
\label{W_conifold}
\end{equation}

\begin{figure}[h]
\begin{center}
\includegraphics[width=9cm]{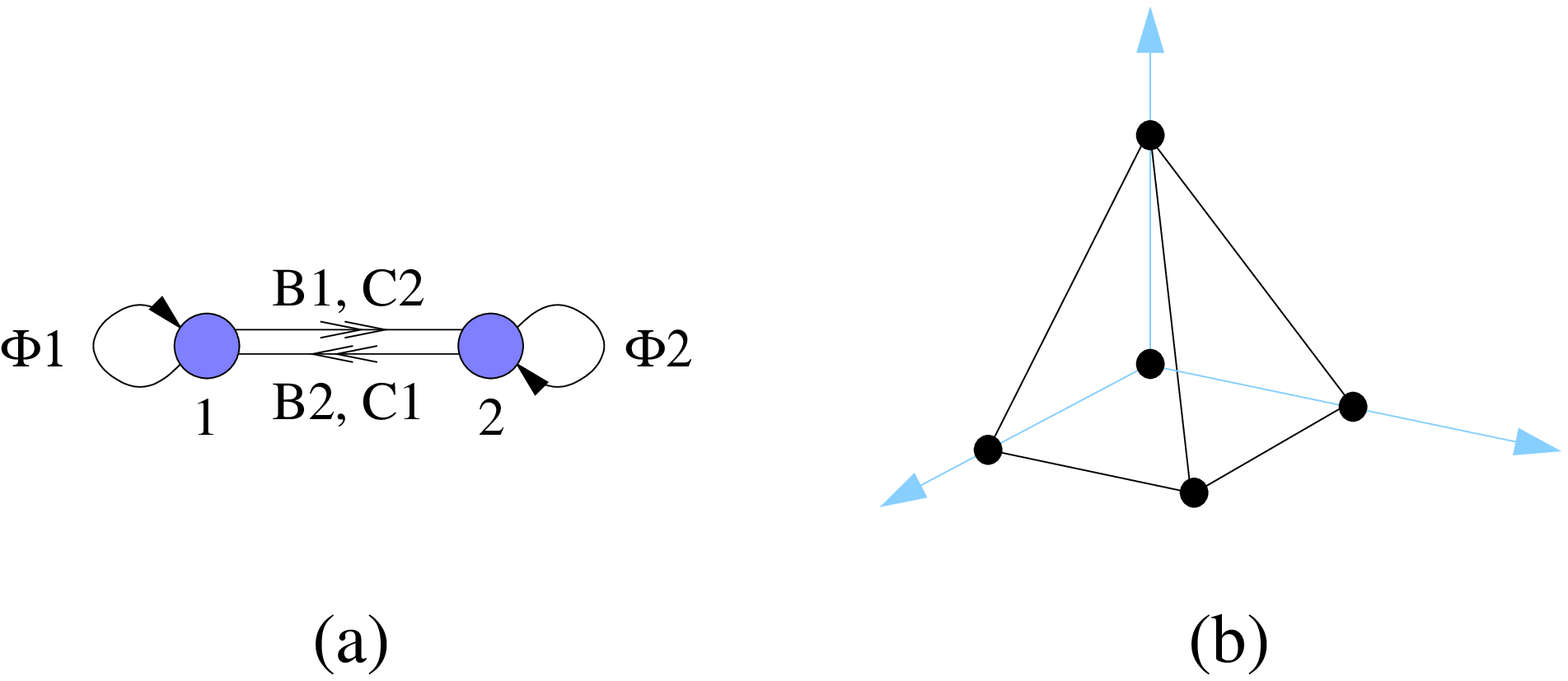}
\caption{Quiver and toric diagram for $C(T^{11}) \times \IIC$.}
\label{conifold}
\end{center}
\end{figure}

This theory has been recently discussed in \cite{Ami}. On the other hand, turning on a vev for $B_1$ resolves 
$\IIC^3/(\IIZ_2 \times \IIZ_2) \times \IIC$ down to $C(SPP) \times \IIC$. This is a new gauge theory without a (3+1)d parent.\footnote{Turning on a vev for $A_1$ leads to a theory in which one of the gauge groups has vanishing CS level. Similarly to what happens for some examples in previous sections, formal computation of the moduli space also leads to $C(SPP) \times \IIC$.} Its quiver diagram
is shown in \fref{SPP}.a, $k=(1,-2,1)$ and its superpotential is given by

\begin{equation}
W= \phi_1 (C_1 A_1-A_2 C_2 B_2) +\phi_2 (B_2 A_2 C_2 -C_1 A_1) ~.
\label{W_SPP}
\end{equation}

\begin{figure}[h]
\begin{center}
\includegraphics[width=9cm]{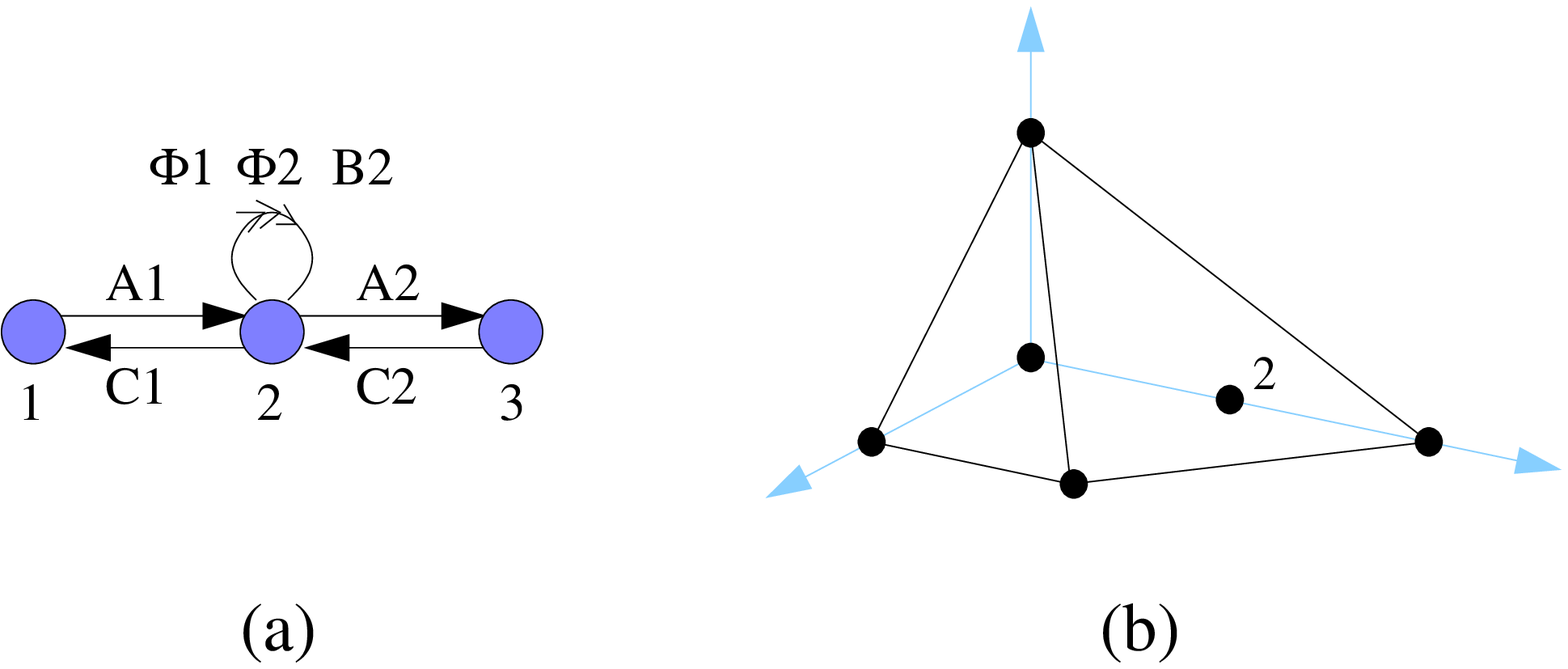}
\caption{Quiver and toric diagram for $C(SPP) \times \IIC$.}
\label{SPP}
\end{center}
\end{figure}

This is in agreement with the crystal proposal \cite{Kim:2007ic}. Computing its moduli space, we obtain the toric diagram in \fref{SPP}.b.

$C(dP_3) \times \IIC$ has a very similar pair of resolutions to the same theories. Vevs for $A_1$ and $A_2$ take us to $C(T^{11}) \times \IIC$, 
and a vev for $B_1$ takes us to $C(SPP) \times \IIC$.

In summary, we have been able to connect all the theories we have discussed in this paper
by either partial resolutions or mass deformations. \fref{roadmap} summarizes the ``roadmap" of connections
between the models.

\begin{figure}[h]
\begin{center}
\psfrag{Z3 x Z3}[cc][][.5]{$\IIC^3/(\IIZ_3 \times \IIZ_3) \times \IIC$}
\psfrag{Z2 x Z2}[cc][][.5]{$\IIC^3/(\IIZ_2 \times \IIZ_2) \times \IIC$}
\psfrag{conifold}[cc][][.6]{$C(T^{11})\times \IIC$}
\psfrag{SPP}[cc][][.6]{$C(SPP)\times \IIC$}
\psfrag{Q111}[cc][][.6]{$C(Q^{111})\times \IIC$}
\psfrag{D3}[cc][][.6]{$D_3$}
\psfrag{dP3}[cc][][.6]{$C(dP_3)\times \IIC$}
\includegraphics[width=10cm]{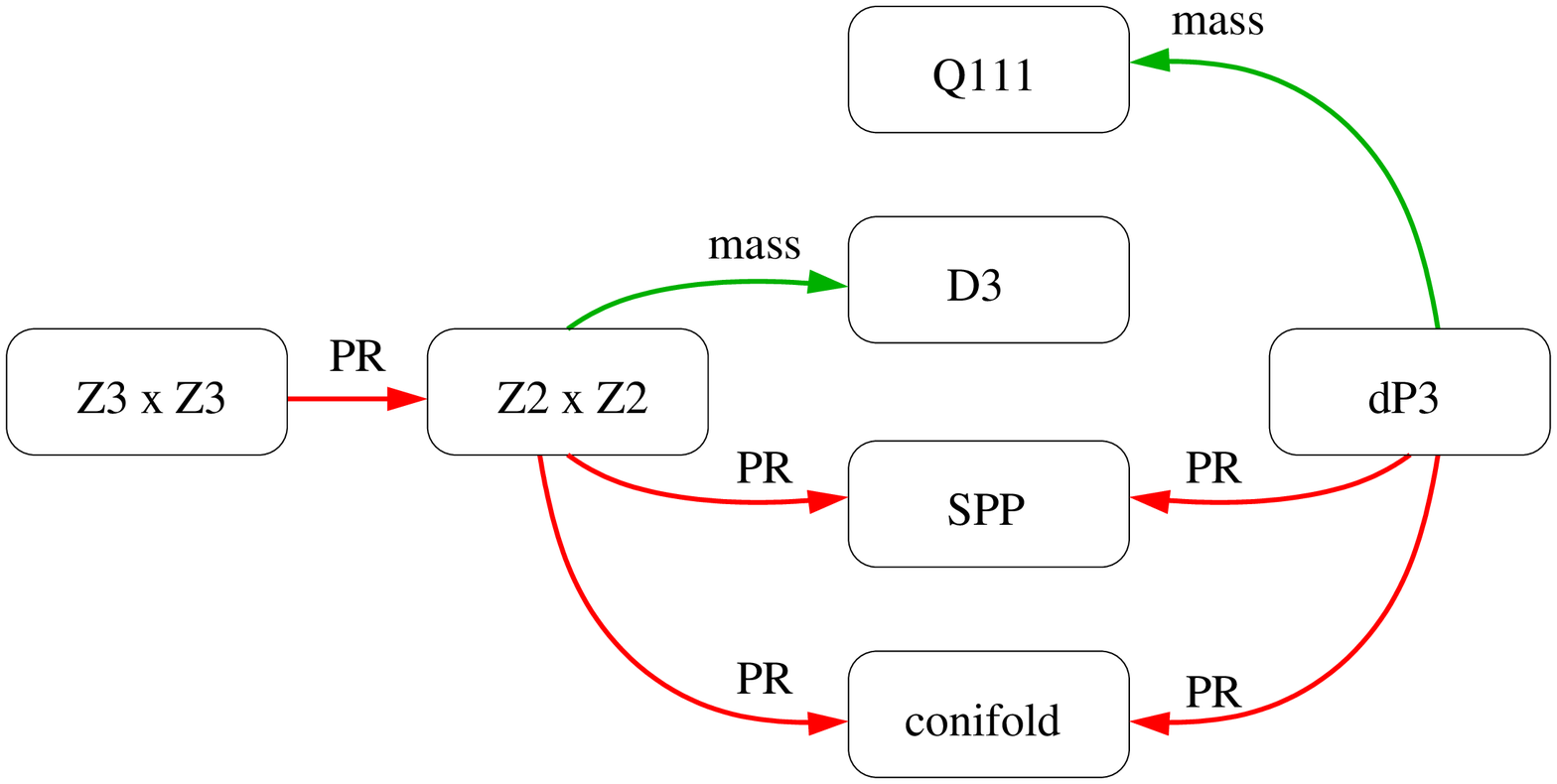}
\caption{Connections between the theories we have studied. PR indicates partial resolution and ``mass" corresponds to RG flows following mass deformations.}
\label{roadmap}
\end{center}
\end{figure}

\section{Conclusions}

\label{section_conclusions}

In this paper we have constructed various examples of (2+1)d ${\cal N}=2$ CS gauge theories that do not have a (3+1)d origin. One of them is the gauge theory for $C(Q^{111})$. We have also considered KW-type RG-flows connecting different theories as well as partial resolutions. It turns out that the chiral field content, superpotentials, RG-flows and partial resolutions are in agreement with crystal models. It is important to emphasize, though, that all our computations, most notably the calculation of moduli spaces, are independent of the validity of crystal models. Thus, our results can be regarded as new evidence that crystal models indeed capture the structure of these theories.

An ambitious goal would be to obtain an efficient procedure for constructing the (2+1)d CS gauge theory for an arbitrary toric CY$_4$-fold, analogous to the one provided by dimer models in (3+1)d. To do this, it is still necessary to understand crystal models in more detail, in particular how they encode gauge groups. The helical path idea of \cite{Hanany:2008fj} seems to be a promising direction. A robust proof of the correspondence between crystal models and CY$_4$/CFT$_3$ is desirable. It is conceivable that the correspondence can be proved both with string theory methods like \cite{Feng:2005gw} or purely in field theoretic terms as in \cite{Franco:2006gc}.

The next step would be to determine all gauge theories whose moduli space is a given geometry. Then, we can investigate whether these models are related by some kind of duality. An interesting Seiberg duality for CS theories has been recently introduced in \cite{Giveon:2008zn}. The full set of dualities might be larger than this since, in general, we expect dual models can have different number of gauge groups. We have briefly mentioned this possibility in section \ref{section_other_pair}, for the case of $D_3$. Interestingly, (2+1)d mirror symmetry is rich in such examples \cite{Intriligator:1996ex}. The $D_3$ models are also examples of theories with and without (3+1)d parents having the same moduli space. A similar pair is the ABJM model and the $N=1$ case of the models in section \ref{section_ZN_ZN}.

Understanding how geometry translates into field theory is the first step towards a general understanding of AdS$_4$/CFT$_3$
in ${\cal N}=2$ settings. In addition, we would like to perform various checks on the dual pairs. One such test, is the precision matching of R-charges computed from field theory and geometry as done in (3+1)d \cite{Hanany1}. At this moment, it is not clear how to implement such program. While the computation can be done on the geometric side using the techniques in \cite{Martelli:2005tp}, it is still not known how to use the field theory ideas of \cite{Barnes:2005bm} in this context.
Another possibility is to work out the BPS operators on both sides of the correspondence, along the lines of \cite{Hanany5}. This program has been already initiated in the context of M2-branes in \cite{Hanany:2008qc,Ami,Hanany:2008fj}. We believe that plenty of new structures are still waiting to be discovered and we hope
to report our progress in the near future.

\centerline{\bf Acknowledgements} \vskip 5mm \noindent

We would like to thank Y.-H. He, C. Herzog, S. Lee, G. Torroba and specially I. Klebanov and A. Zaffaroni for useful discussions.
S.F. is supported by the DOE
under contract DE-FG02-91ER-40671 and by the National Science Foundation under Grant No. PHY05-51164. D. R-G. acknowledges financial support from
the European Commission through Marie Curie OIF grant contract no. MOIF-CT-
2006-38381.
J.P. is  supported in part by the KOSEF SRC Program
  through CQUeST at Sogang University, by KOSEF Grant R01-2008-000-20370-0
  and by  the Stanford Institute for Theoretical Physics.


\appendix
\section{$\IIC^3/(\IIZ_3 \times \IIZ_3) \times \IIC$} \label{A}

In this appendix, we investigate the general proposal of the section \ref{section_ZN_ZN} for the case of $N=3$
The quiver diagram is shown in \fref{quiver_Z3_Z3}. The superpotential is
\beq
W=\phi_1 (X_2 Y_2 Z_2-Z_1 X_1 Y_1)+\phi_2 (X_3 Y_3 Z_3-Z_2 X_2 Y_2)+\phi_3 (X_1 Y_1 Z_1-Z_3 X_3 Y_3) ~.
\eeq

\begin{figure}[h]
\begin{center}
\includegraphics[width=7cm]{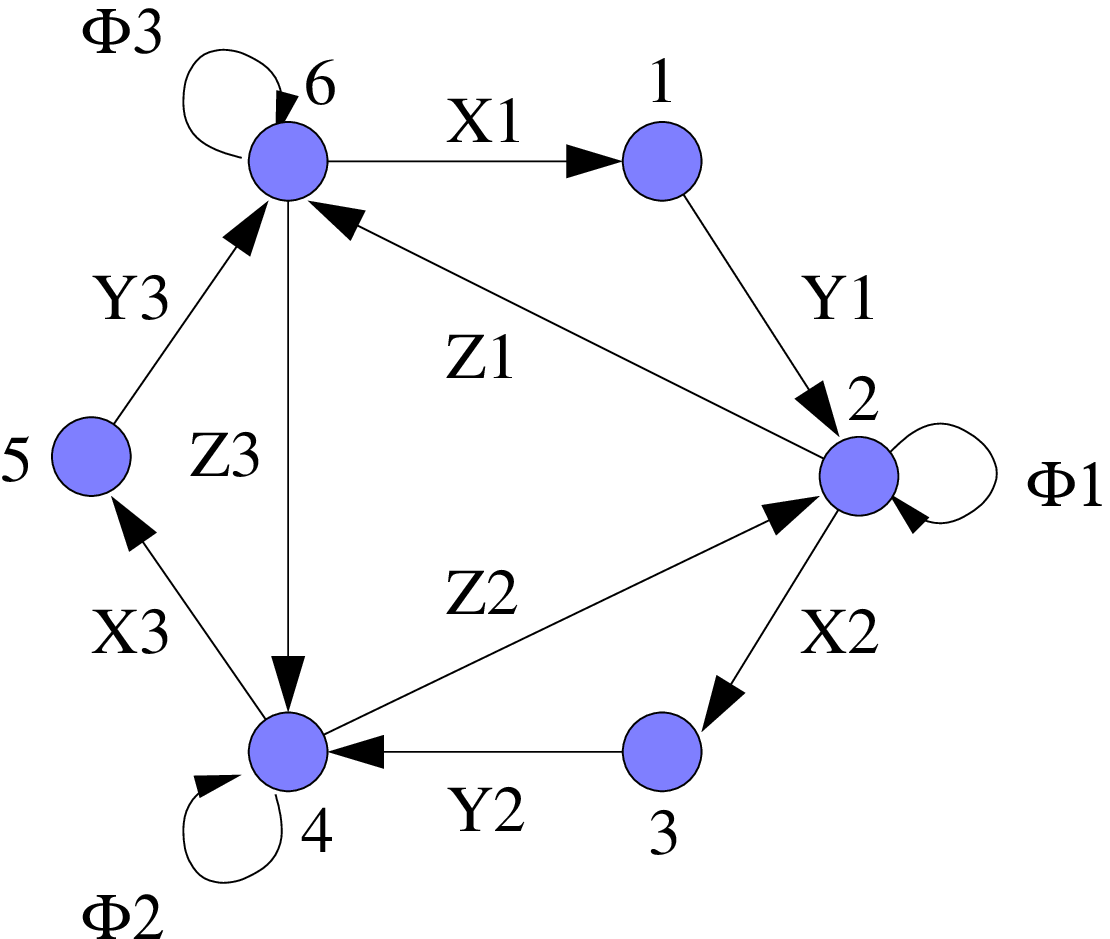}
\caption{Quiver diagram for $\IIC^3/(\IIZ_3 \times \IIZ_3) \times \IIC$.}
\label{quiver_Z3_Z3}
\end{center}
\end{figure}

From this superpotential, we can construct the following Kasteleyn
matrix. Rows and columns 
correspond to negative and positive superpotential terms, respectively
\beq
K=\left(\begin{array}{ccc} 
\phi_1 & 0 & X_1 + Y_1 + Z_1 \\
X_2 + Y_2 + Z_2 & \phi_2 & 0 \\
0 & X_3 + Y_3 + Z_3 & \phi_3
\end{array}\right)
\eeq
The GLSM fields (perfect matchings of the crystal) can be computed as
$\det K$. Notice that although we are 
using technology that is borrowed from the study of dimer models, the
reasoning above is independent of any 
dimer model interpretation and applies to any theory in which the
superpotential satisfies the toric condition 
(i.e. that every field appears in exactly two terms, with opposite
signs). They are 28, and their relation to
 quiver fields is encoded in the following matrix
{\tiny
\beq
P=\left(\begin{array} {c|cccccccccccccccccccccccccccc}
 & p_1 & p_2 & p_3 & p_4 & p_5 & p_6 & p_7 & p_8 & p_9 & p_{10} &
 p_{11} & p_{12} & p_{13} & p_{14} & p_{15} 
& p_{16} & p_{17} & p_{18} & p_{19} & p_{20} & p_{21} & p_{22} & p_{23}& p_{24} & p_{25} & p_{26} & p_{27} & p_{28} \\
X_1 & 1 & 0 & 1 & 0 & 1 & 0 & 1 & 0 & 0 & 0 & 0 & 0 & 1 & 0 & 1 & 0 & 0 & 0 & 1 & 0 & 1 & 0 & 0 & 0 & 1 & 0 & 0 & 0 \\
X_2 & 1 & 1 & 0 & 0 & 1 & 1 & 0 & 0 & 1 & 0 & 1 & 0 & 0 & 0 & 0 & 0 & 0 & 0 & 1 & 1 & 0 & 0 & 1 & 0 & 0 & 0 & 0 & 0 \\
X_3 & 1 & 1 & 1 & 1 & 0 & 0 & 0 & 0 & 1 & 1 & 0 & 0 & 1 & 1 & 0 & 0 & 1 & 0 & 0 & 0 & 0 & 0 & 0 & 0 & 0 & 0 & 0 & 0 \\
Z_1 & 0 & 1 & 0 & 1 & 0 & 1 & 0 & 1 & 0 & 0 & 0 & 0 & 0 & 1 & 0 & 1 & 0 & 0 & 0 & 1 & 0 & 1 & 0 & 0 & 0 & 1 & 0 & 0 \\
Z_2 & 0 & 0 & 1 & 1 & 0 & 0 & 1 & 1 & 0 & 1 & 0 & 1 & 0 & 0 & 0 & 0 & 0 & 0 & 0 & 0 & 1 & 1 & 0 & 1 & 0 & 0 & 0 & 0 \\
Z_3 & 0 & 0 & 0 & 0 & 1 & 1 & 1 & 1 & 0 & 0 & 1 & 1 & 0 & 0 & 1 & 1 & 0 & 1 & 0 & 0 & 0 & 0 & 0 & 0 & 0 & 0 & 0 & 0 \\
Y_1 & 0 & 0 & 0 & 0 & 0 & 0 & 0 & 0 & 1 & 1 & 1 & 1 & 0 & 0 & 0 & 0 & 1 & 1 & 0 & 0 & 0 & 0 & 1 & 1 & 0 & 0 & 1 & 0 \\
Y_2 & 0 & 0 & 0 & 0 & 0 & 0 & 0 & 0 & 0 & 0 & 0 & 0 & 1 & 1 & 1 & 1 & 1 & 1 & 0 & 0 & 0 & 0 & 0 & 0 & 1 & 1 & 1 & 0 \\
Y_3 & 0 & 0 & 0 & 0 & 0 & 0 & 0 & 0 & 0 & 0 & 0 & 0 & 0 & 0 & 0 & 0 & 0 & 0 & 1 & 1 & 1 & 1 & 1 & 1 & 1 & 1 & 1 & 0 \\
\phi_1 & 0 & 0 & 0 & 0 & 0 & 0 & 0 & 0 & 0 & 0 & 0 & 0 & 0 & 0 & 0 & 0 & 0 & 0 & 0 & 0 & 0 & 0 & 0 & 0 & 0 & 0 & 0 & 1 \\
\phi_2 & 0 & 0 & 0 & 0 & 0 & 0 & 0 & 0 & 0 & 0 & 0 & 0 & 0 & 0 & 0 & 0 & 0 & 0 & 0 & 0 & 0 & 0 & 0 & 0 & 0 & 0 & 0 & 1 \\
\phi_3 & 
0 & 0 & 0 & 0 & 0 & 0 & 0 & 0 & 0 & 0 & 0 & 0 & 0 & 0 & 0 & 0 & 0 & 0 & 0 & 0 & 0 & 0 & 0 & 0 & 0 & 0 & 0 & 1
\end{array}
\right).
\eeq}
$Q_F$ is the $20 \times 28$ dimensional matrix obtained as $Ker(P)$. We do not exhibit here for space reasons.
The quiver $U(1)$ charges can be reproduced by the following charge matrix
{\tiny
\beq
\begin{array} {c|cccccccccccccccccccccccccccc}
 & p_1 & p_2 & p_3 & p_4 & p_5 & p_6 & p_7 & p_8 & p_9 & p_{10} &
 p_{11} & p_{12} & p_{13} & p_{14} & p_{15} & p_{16} 
& p_{17} & p_{18} & p_{19} & p_{20} & p_{21} & p_{22} & p_{23}& p_{24} & p_{25} & p_{26} & p_{27} & p_{28} \\ \hline
Q_1 & -1 & 0 & 0 & 0 & 0 & 0 & 0 & 0 & 1 & 0 & 0 & 0 & 0 & 0 & 0 & 0 & 0 & 0 & 0 & 0 & 0 & 0 & 0 & 0 & 0 & 0 & 0 & 0 \\ 
Q_2 & 0 & 1 & 0 & 0 & 0 & 0 & 0 & 0 & 0 & -1 & 0 & 0 & 0 & 0 & 0 & 0 & 0 & 0 & 0 & 0 & 0 & 0 & 0 & 0 & 0 & 0 & 0 & 0 \\
Q_3 & 0 & -1 & 0 & 0 & 0 & 0 & 0 & 0 & 0 & 0 & 0 & 0 & 0 & 1 & 0 & 0 & 0 & 0 & 0 & 0 & 0 & 0 & 0 & 0 & 0 & 0 & 0 & 0 \\
Q_4 & 0 & 0 & 1 & 0 & 0 & 0 & 0 & 0 & 0 & 0 & 0 & 0 & 0 & 0 & -1 & 0 & 0 & 0 & 0 & 0 & 0 & 0 & 0 & 0 & 0 & 0 & 0 & 0 \\
Q_5 & 0 & -1 & 0 & 0 & 0 & 0 & 0 & 0 & 0 & 0 & 0 & 0 & 0 & 0 & 0 & 0 & 0 & 0 & 0 & 1 & 0 & 0 & 0 & 0 & 0 & 0 & 0 & 0 \\
Q_6 & 1 & 1 & -1 & 0 & 0 & 0 & 0 & 0 & -1 & 1 & 0 & 0 & 0 & -1 & 1 & 0 & 0 & 0 & 0 & -1 & 0 & 0 & 0 & 0 & 0 & 0 & 0 & 0 
\end{array}
\eeq}
Following the general proposal, we take CS levels
$k=(1,-1,1,-1,1,-1)$. We then quotient by $Q_1+Q_2$, $Q_1+Q_4$, 
$Q_1+Q_6$ and $Q_2+Q_3$. Then, we have
{\tiny
\beq
Q_D=\left( \begin{array} {cccccccccccccccccccccccccccc}

p_1 & p_2 & p_3 & p_4 & p_5 & p_6 & p_7 & p_8 & p_9 & p_{10} & p_{11}
& p_{12} & p_{13} & p_{14} & p_{15} & p_{16} & 
p_{17} & p_{18} & p_{19} & p_{20} & p_{21} & p_{22} & p_{23}& p_{24} & p_{25} & p_{26} & p_{27} & p_{28} \\ \hline
-1 & 1 & 0 & 0 & 0 & 0 & 0 & 0 & 1 & -1 & 0 & 0 & 0 & 0 & 0 & 0 & 0 & 0 & 0 & 0 & 0 & 0 & 0 & 0 & 0 & 0 & 0 & 0 \\ 
-1 & 0 & 1 & 0 & 0 & 0 & 0 & 0 & 1 & 0 & 0 & 0 & 0 & 0 & -1 & 0 & 0 & 0 & 0 & 0 & 0 & 0 & 0 & 0 & 0 & 0 & 0 & 0 \\
0 & 1 & -1 & 0 & 0 & 0 & 0 & 0 & 0 & 1 & 0 & 0 & 0 & -1 & 1 & 0 & 0 & 0 & 0 & -1 & 0 & 0 & 0 & 0 & 0 & 0 & 0 & 0 \\
0 & 0 & 0 & 0 & 0 & 0 & 0 & 0 & 0 & -1 & 0 & 0 & 0 & 1 & 0 & 0 & 0 & 0 & 0 & 0 & 0 & 0 & 0 & 0 & 0 & 0 & 0 & 0 
\end{array} \right)
\eeq}
We combine $Q_F$ and $Q_D$ into $Q_{tot}$ and calculate the toric
diagram of the moduli space as $G^T=Ker(Q_{tot})$.  The result is
{\tiny
\beq
G^T=\left( \begin{array} {cccccccccccccccccccccccccccc}

p_1 & p_2 & p_3 & p_4 & p_5 & p_6 & p_7 & p_8 & p_9 & p_{10} & p_{11}
& p_{12} & p_{13} & p_{14} & p_{15} & p_{16} & 
p_{17} & p_{18} & p_{19} & p_{20} & p_{21} & p_{22} & p_{23}& p_{24} & p_{25} & p_{26} & p_{27} & p_{28} \\ \hline
0 & 0 & 0 & 0 & 0 & 0 & 0 & 0 & 0 & 0 & 0 & 0 & 0 & 0 & 0 & 0 & 0 & 0 & 0 & 0 & 0 & 0 & 0 & 0 & 0 & 0 & 0 & 1 \\
-2 & -2 & -2 & -2 & -2 & -2 & -2 & -2 & -1 & -1 & -1 & -1 & -1 & -1 & -1 & -1 & 0 & 0 & -1 & -1 & -1 & -1 & 0 & 0 & 0 & 0 & 1 & 0 \\
0 & 1 & 1 & 2 & 1 & 2 & 2 & 3 & 0 & 1 & 1 & 2 & 0 & 1 & 1 & 2 & 0 & 1 & 0 & 1 & 1 & 2 & 0 & 1 & 0 & 1 & 0 & 0 \\
3 & 2 & 2 & 1 & 2 & 1 & 1 & 0 & 2 & 1 & 1 & 0 & 2 & 1 & 1 & 0 & 1 & 0 & 2 & 1 & 1 & 0 & 1 & 0 & 1 & 0 & 0 & 0 
\end{array} \right)
\eeq}
In \fref{toric_Z3_Z3} we plot the first three rows of this matrix. This corresponds precisely to the toric diagram of $\IIC^3/(\IIZ_3 \times \IIZ_3) \times \IIC$ and has a nice structure of multiplicities.

\begin{figure}[h]
\begin{center}
\includegraphics[width=6cm]{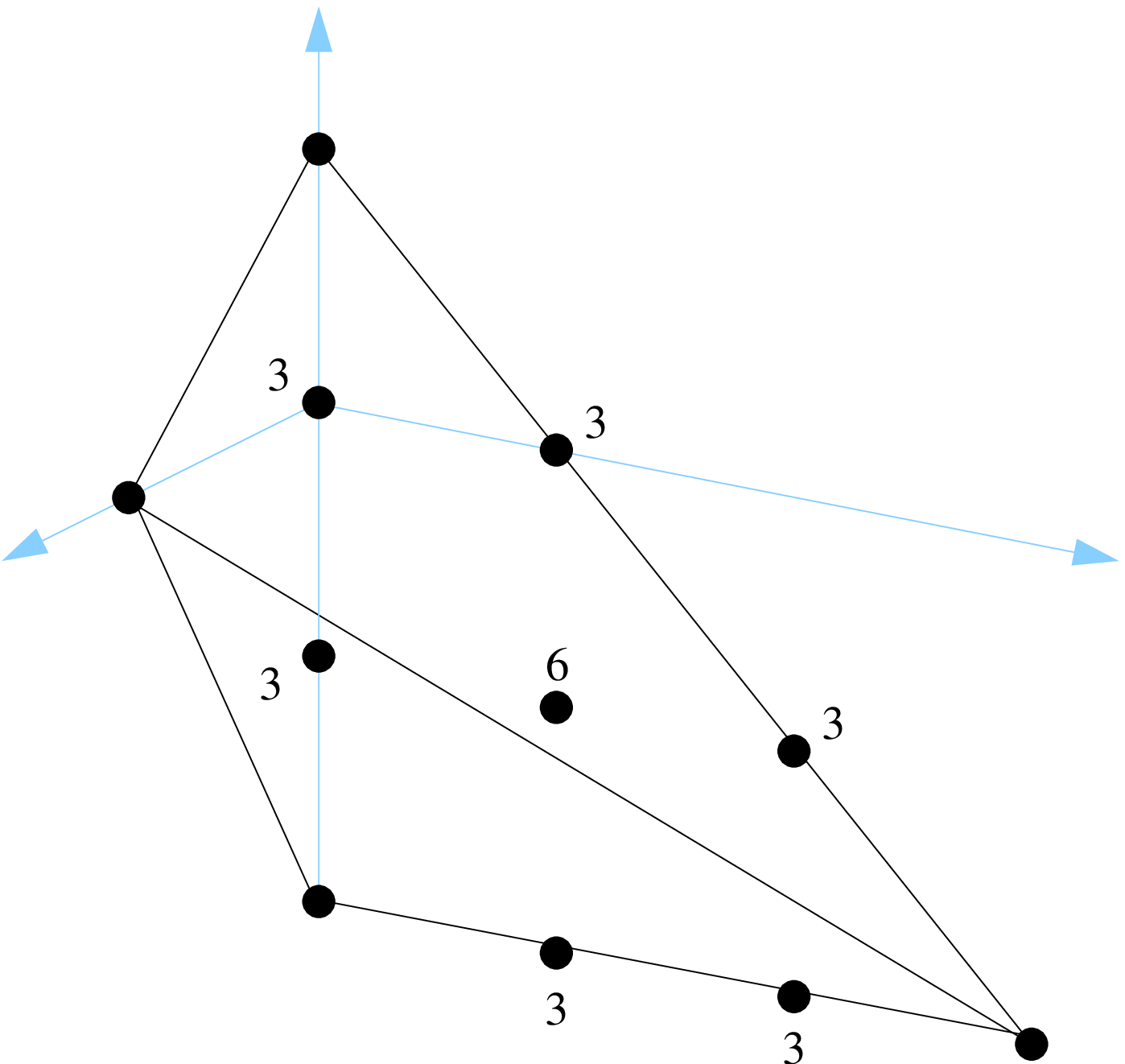}
\caption{Toric diagram for the $k=(1,-1,1,-1,1,-1)$ theory.}
\label{toric_Z3_Z3}
\end{center}
\end{figure}

\section{Parity invariance}

\label{section_parity}

Parity invariance is a key property expected to be satisfied by M2-brane theories.
In this appendix we present some evidence that our models preserve parity invariance. More concretely, we show that when we expand the action around a point in moduli space at which gauge groups with opposite CS levels are higgsed to the diagonal subgroup, parity invariance is preserved up to irrelevant terms (for some assumption about the superpotential). Our method is similar to the one used in \cite{Mukhi:2008ux} to derive the action of D2-branes from that of M2-branes.

In order to illustrate our strategy, let us consider the toy model shown in \fref{quiver_toy_parity}. The gauge group is $U(N_1) \times U(N_2)$ and we have two bifundamentals $X_{12}$, $X_{21}$
and one adjoint $\Phi$ for the second group. This theory contains various structures that are present in general models.

\begin{figure}[h]
\begin{center}
\psfrag{X12}[cc][][1]{$X_{12}$}
\psfrag{X21}[cc][][1]{$X_{21}$}
\includegraphics[width=4cm]{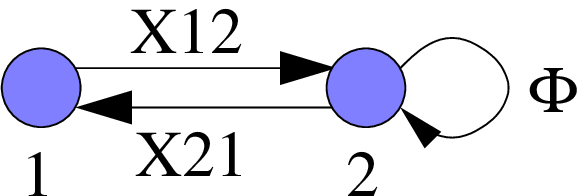}
\caption{A toy model we use to investigate our ideas about parity invariance.}
\label{quiver_toy_parity}
\end{center}
\end{figure}

The action is given by 
\begin{eqnarray}
S&=&\frac{1}{2} \int d^3 x (-D_{\m}X_{12} D^{\m} X_{12}^{\dagger}-D_{\m}X_{21}
D^{\m} X_{21}^{\dagger}-D_{\m}\Phi D^{\m} \Phi ) \nn \\
 &+ & \frac{k}{2}\int d^3 x \epsilon^{\m\n\l}
 ({\cal A}^1_{\m}\partial_{\n}{\cal A}^1_{\lambda}+\frac{2i}{3}{\cal A}^1_{\m}{\cal A}^1_{\n}{\cal A}^1_{\l}
-{\cal A}^2_{\m}\partial_{\n}{\cal A}^2_{\lambda}-\frac{2i}{3}{\cal A}^2_{\m}{\cal A}^2_{\n}{\cal A}^2_{\l}) + \ldots
\end{eqnarray}
where the covariant derivatives are

\begin{eqnarray}
D_{\m} X_{12}&=&\partial_{\m}X_1+i({\cal A}^1_{\m} X_{12}-X_{12}{\cal A}^2_{\m}) \nn \\
D_{\m} X_{21}&=&\partial_{\m}X_2+i({\cal A}^2_{\m} X_{21}-X_{21}{\cal A}^1_{\m}) \nn \\
D_{\m}\Phi &=& \partial_{\m} \Phi+i[{\cal A}^2_{\m}, \Phi] 
\end{eqnarray}
with $\Phi$ being Hermitian. We leave traces implicit in all our expressions. Next, let us define the combinations 

\beq
2 {\cal A}^{\pm} = {\cal A}^1 \pm {\cal A}^2 ~.
\eeq

The CS term can be rewritten as follows 
\begin{eqnarray}
\epsilon^{\m\n\l}({\cal A}^1_{\m}\partial_{\n}{\cal A}^1_{\l}-{\cal A}^2_{\m}\partial_{\n}{\cal A}^2_{\l})
&=& 4 \epsilon^{\m\n\l} {\cal A}^{-}_{\m} \partial_{\nu} {\cal A}^{+}_{\l} \nn \\
\epsilon^{\m\n\l}({\cal A}^1_{\m}{\cal A}^1_{\n}{\cal A}^1_{\l}-{\cal A}^2_{\m}{\cal A}^2_{\n}{\cal A}^2_{\l})
&=& 2 \epsilon^{\m\n\l} ({\cal A}^{-}_{\m} {\cal A}^{-}_{\n} {\cal A}^{-}_{\l}+3 {\cal A}^{-}_{\m} {\cal A}^{+}_{\n} {\cal A}^{+}_{\l}) ~.
\end{eqnarray}
We can also write 
\beq
\begin{array}{rcl}
D_{\m} X_{12} & = & \partial_{\m} X_1+i[{\cal A}^{+}, X_{12}]+i({\cal A}^{-}_{\m}
X_{12}+X_{12} {\cal A}^{-}_{\m}) \\
 & \equiv  & D^+_{\m} X_{12}+i({\cal A}^{-}_{\m} X_{12}+X_{12} {\cal A}^{-}_{\m}) \\ \\
D_{\m} X_{12}^{\dagger}& = & \partial_{\m} X_{12}^{\dagger}+i[{\cal A}^{+}, X_{12}^{\dagger}]+i({\cal A}^{-}_{\m}
X_{12}^{\dagger}+X_{12}^{\dagger}{\cal A}^{-}_{\m}) \\ 
& \equiv & D^+_{\m} X_{12}^{\dagger}
+i({\cal A}^{-}_{\m} X_{12}^{\dagger}+X_{12}^{\dagger} {\cal A}^{-}_{\m}) 
\end{array}
\eeq

Next, let us expand around some point in moduli space $X_{12}=R \, {\bf 1}_{N \times N}$ such that  $X_{12}=R \, {\bf 1}_{N \times
  N}+ \tilde{X}_{12}$.

\begin{equation}
\begin{array}{rcl}
D_{\m}X_{12} D^{\m} X_{12}^{\dagger} & = & D^+_{\m}\tilde{X}_{12}
D^{+\m}\tilde{X}_{12}^{\dagger}+{\cal A}^{-}_{\m}{\cal A}^{-\m}(R^2+{\cal O}(\tilde{X}_{12}^2)) \\
& + & i(2R {\cal A}^{-}_{\m}+{\cal A}^{-}_{\m}\tilde{X}_{12}+\tilde{X}_{12} {\cal A}^{-}_{\m})D^{+\m} \tilde{X}_{12}^{\dagger} \\
& - & i(2R{\cal A}^{-}_{\m}+{\cal A}^{-}_{\m} \tilde{X}_{12}^{\dagger}+\tilde{X}_{12}^{\dagger}{\cal A}^{-}_{\m})D^{+\m}\tilde{X}_{12} \\ \\

D_{\m}X_{21} D^{\m} X_{21}^{\dagger} & = & D^+_{\m}X_{21}D^{+\m}X_{21}^{\dagger}
+ i({\cal A}^{-}_{\m}X_{21}+X_{21} {\cal A}^{-}_{\m})D^{+\m} X_{21}^{\dagger} \\
& - & i({\cal A}^{-}_{\m} X_{21}^{\dagger}+X_{21}^{\dagger}{\cal A}^{-}_{\m})D^{+\m} X_{21}  \\ \\

(D_{\m}\Phi)^2 & = & (D^+_{\m}\Phi)^2-2i[{\cal A}^{-}_{\m}, \Phi]D^+_{\m}\Phi-([{\cal A}^{-}_{\m}, \Phi])^2 
\end{array}
\end{equation}

The action does not contain any derivative of ${\cal A}^{-}_\m$. Then, similarly to \cite{Mukhi:2008ux}, we can eliminate it from the action using its equation of motion, resulting in

\begin{eqnarray}
S&=& \frac{1}{2} \int d^3 x (-D^+_{\m} \tilde{X}_{12} D^{+\m} \tilde{X}_{12}^{\dagger}-D^+_{\m}
X_{21} D^{+\m} X_{21}^{\dagger}-D^+_{\m} \Phi D^{+\m} \Phi \nn \\
& &\frac{1}{4(R^2+{\cal O}(\tilde{X}_{12}^2,X_{21}^2, \Phi^2))} G_{\m}G^{\m} +i {\cal A}^{-}_{\m} {\cal A}^{-}_{\n}
{\cal A}^{-}_{\l}) + \ldots
\end{eqnarray}
We are indeed integrating out ${\cal A}^{-}_\m$. The last term should be understood as a shorthand for what results from replacing ${\cal A}^{-}$ by the equation of motion. We have also defined

\begin{eqnarray}
G^{\m}&=&
4k\epsilon^{\m\n\l}D^+_{\n}{\cal A}^{+}_{\l}+2iRD^{+\m} X_{12}^{\dagger}-2iR D^{+\m}X_{12}
\nn \\
&+& \sum_{i=\{12,21\}} (2i X_i D^{+\m}X_i^{\dagger}-2i X_i^{\dagger}
D^{\m}X_i+2iD^{+\m}X_i^{\dagger}X_i
-2iD^{+\m}X_i X_i^{\dagger} \nn \\
& & -2i[, \Phi]D^{+\m}\Phi) ~.
\label{G}
\end{eqnarray}
Starting from the previous equation, we drop the tilde in $\tilde{X}_{12}$.
The $D^+_{\n}{\cal A}^{+}_{\l}$ squared term gives the usual YM kinetic term. The commutator term in the last line of \eref{G}
comes from $[{\cal A}^{-}_{\m}, \Phi]D^{+\m}\Phi$ in the action, from which we have extracted ${\cal A}^{-}_{\m}$. Componentwise,
the last line involves the structure constants $f^{abc}$ of the Lie algebra. $R$ plays the role of a perturbation expansion parameter. If the superpotential is quartic (say with terms of the form $X_{12}X_{21}\Phi^2$) ${\cal A}^{+}_{\l}$, $X_i$ and $\Phi$ have canonical dimension $1/2$.

Parity acts by, for example, $x_1 \rightarrow -x_1$. We can make \eref{G} invariant if $X_i \rightarrow X_i^{\dagger}$ and $\Phi$ does not change under a parity transformation. Notice that $X_i \rightarrow
X_i^{\dagger}$ is the same type of transformation used in ABJM to achieve parity invariance \cite{ABJM}. In ABJM, this operation is accompanied by exchanging the two gauge groups. In our notation, flipping the gauge groups corresponds to ${\cal A}^{-}_{\m} \to -{\cal A}^{-}_{\m}$. Since we have integrated out ${\cal A}^{-}_{\m}$, this last transformation is not visible in our formalism.

Terms involving $\Phi$ are irrelevant. So is the $({\cal A}^{-})^3$ term after using the equation of motion. Thus, the parity violating 
terms vanish in the IR limit. We expect this kind of argument can be applied to generic points in moduli space. We can regard the procedure we have just outlined as going to some kind of unitary gauge. The method is a bit subtle, since the transformation is singular when $(R^2+{\cal O}(X_i^2, \Phi^2))$ vanishes. Our arguments are based on the chiral fields having dimension $1/2$. This issue becomes more subtle for sextic superpotentials, but we have already seen that models with sextic superpotential such as $C(Q^{111})$ can be regarded as models with a quartic
superpotential by adding massive adjoints.

This method can be applied to most of the models in our paper, in which we can separate gauge groups into pairs with $(k,-k)$ CS levels.\footnote{The model in \fref{SPP} has $k=(1,-2,1)$ and does not fall into this category. It would be interesting to study how our ideas extend to more general models.} Let us, for example, consider the $C(Q^{111})$ theory. We have

\begin{eqnarray}
S&=& -\frac{1}{2} \int (D_{\m}A_iD^{\m}A_i^{\dagger}+
D_{\m}B_iD^{\m}B_i^{\dagger}+ D_{\m}C_iD^{\m}C_i^{\dagger}) \nn \\
&+ &\frac{k}{2} \int \epsilon^{\m\n\l}({\cal A}_1\partial_{\n} {\cal A}_1-{\cal A}_3\partial_{\n} {\cal A}_3
+{\cal A}_2\partial_{\n} {\cal A}_2-{\cal A}_4\partial_{\n} {\cal A}_4)+ \cdots
\end{eqnarray}
and 
\begin{eqnarray}
D_{\m}A_1&=&\partial_{\m} A_1+i({\cal A}_1-{\cal A}_3)A_1 \nn \\
D_{\m}C_2&=&\partial_{\m} C_2+i({\cal A}_4-{\cal A}_2)C_2 ~.
\end{eqnarray}
Proceeding as before, we can rewrite the CS term as 
\begin{equation}
\epsilon^{\m\n\l}(({\cal A}_1-{\cal A}_3)_{\m}\partial_{\n}({\cal A}_1+{\cal A}_3)_{\l}
+({\cal A}_2-{\cal A}_4)_{\m}\partial_{\n}({\cal A}_2+{\cal A}_4)_{\l}).
\end{equation}
Expanding around $A_1=R$ and $C_2=r$, we can integrate out the $({\cal A}_1-{\cal A}_3)$ and
$({\cal A}_2-{\cal A}_4)$ combinations. Then, we see parity invariance can be achieved by $A_i, B_i, C_i
\rightarrow A_i^{\dagger}, B_i^{\dagger}, C_i^{\dagger}$ up to
irrelevant terms.


\end{document}

%% file: texpref.tex
\newcommand{\eref}[1]{(\ref{#1})}

\newcommand{\fref}[1]{Figure~\ref{#1}}
\newcommand{\cref}[1]{Chapter~\ref{#1}}
\newcommand{\beq}{\begin{equation}}
\newcommand{\eeq}{\end{equation}}
\newcommand{\ba}{\begin{array}}
\newcommand{\ea}{\end{array}}
\newcommand{\bcenter}{\begin{center}}
\newcommand{\ecenter}{\end{center}}

\def\IB{\relax\hbox{$\inbar\kern-.3em{\rm B}$}}
\def\IC{\relax\hbox{$\inbar\kern-.3em{\rm C}$}}
\def\ID{\relax\hbox{$\inbar\kern-.3em{\rm D}$}}
\def\IE{\relax\hbox{$\inbar\kern-.3em{\rm E}$}}
\def\IF{\relax\hbox{$\inbar\kern-.3em{\rm F}$}}
\def\IG{\relax\hbox{$\inbar\kern-.3em{\rm G}$}}
\def\IGa{\relax\hbox{${\rm I}\kern-.18em\Gamma$}}
\def\IH{\relax{\rm I\kern-.18em H}}
\def\IK{\relax{\rm I\kern-.18em K}}
\def\IL{\relax{\rm I\kern-.18em L}}
\def\IP{\relax{\rm I\kern-.18em P}}
\def\IR{\relax{\rm I\kern-.18em R}}
\def\IZ{\relax\ifmmode\mathchoice
{\hbox{\cmss Z\kern-.4em Z}}{\hbox{\cmss Z\kern-.4em Z}}
{\lower.9pt\hbox{\cmsss Z\kern-.4em Z}}
{\lower1.2pt\hbox{\cmsss Z\kern-.4em Z}}\else{\cmss Z\kern-.4em Z}\fi}
\def\II{\relax{\rm I\kern-.18em I}}

\def\sCC{{\kern 0.27em\vrule height1.45ex width0.03em depth0em
          \kern-0.30em\rm C}}
\def\C{{\mathchoice
  {\sCC}
  {\sCC}
  {\kern 0.225em \vrule height1.05ex width0.025em depth0em \kern-0.25em \rm C}
  {\kern 0.180em \vrule height0.78ex width0.02em depth0em \kern-0.2em \rm C}
        }}
\def\sHH{{\rm I\kern-.16em{}H}}
\def\H{{\mathchoice
  {\sHH}
  {\sHH}
  {\rm I\kern-.13em{}H}
  {\rm I\kern-.13em{}H} }}
\def\sNN{{\rm I\kern-.16em{}N}}
\def\N{{\mathchoice
  {\sNN}
  {\sNN}
  {\rm I\kern-.12em{}N}
  {\rm I\kern-.10em{}N} }}
\def\sPP{{\rm I\kern-.16em{}P}}
\def\P{{\mathchoice
  {\sPP}
  {\sPP}
  {\rm I\kern-.12em{}P}
  {\rm I\kern-.10em{}P} }}
\def\sQQ{{\kern 0.27em \vrule height1.45ex width0.03em depth0em
          \kern-0.30em \rm Q}}
\def\Q{{\mathchoice
        {\sQQ}
        {\sQQ}
  {\kern 0.225em \vrule height1.05ex width0.025em depth0em \kern-0.25em \rm Q}
  {\kern 0.180em \vrule height0.78ex width0.020em depth0em \kern-0.20em \rm Q}
        }}
\def\sRR{{\rm I\kern-0.16em{}R}}
\def\R{{\mathchoice
  {\sRR}
  {\sRR}
  {\rm I\kern-0.12em{}R}
  {\rm I\kern-0.10em{}R} }}
\def\sZZ{{\rm Z\kern-0.32em{}Z}}
\def\Z{{\mathchoice
  {\sZZ}
  {\sZZ} 
  {\rm Z\kern-0.3em{}Z}     
  {\rm Z\kern-0.25em{}Z} }}  
\def\ZZZ{{\rm Z\kern-0.24em{}Z}}
\def\sII{{\rm I\kern-0.16em{}I}}
\def\I{{\mathchoice
  {\sII}
  {\sII}
  {\rm I\kern-0.12em{}I}
  {\rm I\kern-0.10em{}I} }}

\def\inbar{\,\vrule height1.5ex width.4pt depth0pt}
\font\cmss=cmss10 \font\cmsss=cmss10 at 7pt

\def\smiley{\hbox{\large$\bigcirc$\hspace{-0.80em}\raise.2ex
\hbox{$\cdot\cdot$}\kern-.61em\lower.2ex\hbox{\scriptsize$\smile$}}\ }
\def\frowny{\hbox{\large$\bigcirc$\hspace{-0.80em}\raise.2ex
\hbox{$\cdot\cdot$}\kern-.635em\lower.2ex\hbox{\scriptsize$\frown$}}\ }

\def\I{{\rlap{1} \hskip 1.6pt \hbox{1}}}

\makeatletter
\let\hangafter\@hangfrom
\makeatother

\newcommand{\drawsquare}[2]{\hbox{%
\rule{#2pt}{#1pt}\hskip-#2pt
\rule{#1pt}{#2pt}\hskip-#1pt
\rule[#1pt]{#1pt}{#2pt}}\rule[#1pt]{#2pt}{#2pt}\hskip-#2pt
\rule{#2pt}{#1pt}}

\newcommand{\fund}{\raisebox{-.5pt}{\drawsquare{6.5}{0.4}}}
\newcommand{\antifund}{\overline{\fund}}